\documentclass[aps,prb,twocolumn,superscriptaddress,nofootinbib]{revtex4-2}

\usepackage{braket}
\usepackage{float}
\usepackage{comment}
\usepackage{amsmath,amssymb,bm,graphicx}
\usepackage{hyperref}
\newcommand{\Tr}{\mbox{Tr }}

\begin{document}

\title{Signatures of slow particle transport in the interacting Aubry-Andr\'e chain}

\author{Sara Sharifan}
\affiliation{Department of Physics and Astronomy, University of Manitoba, Winnipeg, Manitoba R3T 2N2, Canada}

\author{Maximilian Kiefer-Emmanouilidis}
\affiliation{Theoretical Physics,  Saarland University,  D-66123  Saarbr\"ucken,  Germany}
\affiliation{Center for Quantum Technologies (QuTe), Saarland University, D-66123 Saarbr\"ucken, Germany}
\affiliation{Embedded Intelligence, German Research Center for Artificial Intelligence (DFKI), D-67663 Kaiserslautern, Germany}
\affiliation{Department of Computer Science and Research Initiative QC-AI, RPTU Kaiserslautern-Landau, D-67663 Kaiserslautern, Germany}
%\affiliation{Department of Physics, RPTU Kaiserslautern-Landau, Kaiserslautern, Germany}

\author{Jesko Sirker}
\affiliation{Department of Physics and Astronomy, University of Manitoba, Winnipeg, Manitoba R3T 2N2, Canada}

\begin{abstract}
Slow delocalization in strongly disordered many-body quantum chains can originate from two distinct mechanisms: rare resonances and avalanches, rooted in the energy statistics, or an energy-independent combinatorial path-entropy obstruction in operator space. Quasiperiodic potentials eliminate rare regions with similar potentials that are thought to be responsible for avalanche instabilities and modify the statistics of resonances, thus allowing us to delineate the two mechanisms. To do so, we numerically study the entanglement entropy $S(t)$, number entropy $S_N(t)$, and Hartley number entropy $S_H(t)$ following a quantum quench in the interacting Aubry-Andr\'e chain with disorder strength $D$. We find $S(t)\sim\mu\ln t$, $S_N(t)\sim\nu\ln\ln t$ with growth rates $\mu\sim\nu\sim D^{-\gamma}$ and $\gamma\approx 3$, the same behavior that was previously found in the case of quenched disorder. Furthermore, the saturation values $S_H(t\to\infty)$ continue to grow with system size $L$ as $\ln L$, a growth that persists for $D\gtrsim 8$ where MBL was argued to be stable. The similarity of the quasi-periodic and quenched disorder cases points to a common origin of the slow particle transport that does not rely on rare-region avalanches and persists despite strongly modified resonance statistics.
\end{abstract}

\maketitle

\section{Introduction}
Localization has been a cornerstone of disordered quantum physics since Anderson's work on the absence of diffusion in random lattices \cite{Anderson58}. In 2006, Basko, Aleiner, and Altshuler argued that localization might survive even in the presence of interactions \cite{AleinerAltshuler}. This gave rise to the concept of many-body localization (MBL), thought of as representing a breakdown of thermalization in interacting systems \cite{NandkishoreHuse,AbaninRev2019}. Studies of the MBL phase and its associated ergodic to non-ergodic phase transition have been largely based on numerical studies of small systems \cite{ZnidaricProsen,PalHuse,OganesyanHuse,Luitz1,Luitz2}. By 2019, a broad consensus had formed that MBL can indeed be a phase stable in the thermodynamic limit in one dimensional systems with quenched disorder. This consensus was, however, then questioned based on the observations of (1) an apparent shift in the critical disorder strength with increasing system size \cite{SuntajsBonca,SuntajsBonca2}, (2) an analysis of the fidelity scaling \cite{SelsPolkovnikov,SelsPolkovnikov2,SelsPolkolnikovReview}, and (3) an analysis of the number and Hartley number entropies \cite{KieferSciPost2020,KieferUnanyan2,KieferUnanyan3,KieferUnanyan4,KieferSciPost2022}.

On the much less explored analytical side, efforts have been focussed on investigating if a microscopic Hamiltonian in an MBL phase can be transformed unitarily into an effective Hamiltonian of quasi-local charges \cite{Imbrie2016,Imbrie_JSTAT,ImbrieRosScardicchio,deRoeckGiacomin,huveneersMBL}. Here, two distinct obstructions occur. On the one hand, many-body resonances and so-called avalanche instabilities---caused by rare regions with small potential differences---have to be kept under control \cite{Imbrie_JSTAT,MorningstarHuse,DeRoeckHuveneers}. On the other hand, there is a potential path-entropy instability in many-body systems which is independent of the distribution of energies. While in a Gaussian system, there are only $\sim N^2$ possible operators for a system of size $N$ and the commutator branching is $\mathcal{O}(1)$, there are exponentially many possible operators in an interacting system and the branching is extensive \cite{WeisseGerstnerSirker,SirkerObstruction}.  

The goal of this work is to try to obtain evidence to distinguish between these two potential instabilities by studying the quasi-periodic interacting Aubry-Andr\'e chain. This allows us to define potentials such that the potential energies near the bond where the system is cut in half are exactly the same for different system sizes. Only the sites added at the two ends of the chains when increasing the system size have novel potentials. Furthermore, we choose Diophantine frequencies such that extended regions of almost equal potential are impossible and near-repetitions of the potential landscape only occur at large distances. Avalanche instabilities therefore play no role by construction and the statistics of pairwise resonances is strongly modified as compared to the quenched disorder case. If many-body resonances and avalanches are the main factors in destabilizing the potential MBL phase, then one would therefore expect that the results for the Aubry-Andr\'e chain are qualitatively different from the case of quenched disorder. 

To address this question, we will use exact diagonalizations and a Trotterization of the time-evolution operator to study the dynamics after a quantum quench. Specifically, we will calculate the entanglement entropy $S(t)$, number entropy $S_N(t)$, and Hartley number entropy $S_H(t)$ for cutting a system with an even number of sites into two equal halves. For the entanglement entropy, it has been suggested that the change from a power-law growth, $S(t)\sim t^\alpha$, to a slower logarithmic growth $S(t)\sim\ln t$, is a signature of the ergodic-MBL phase transition \cite{ZnidaricProsen,BardarsonPollmann,Serbyn2013}. For systems where the total particle number is conserved, like the Aubry-Andr\'e chain studied here, the reduced density matrix $\rho_A$ of the subsystem $A$ has a block structure with respect to the number of particles $N_A$ in the subsystem. This block structure can be used to split the entanglement entropy into two components $S=S_C+S_N$ with \cite{MonkmanSirker4,LukinGreinerMBL}
\begin{eqnarray}
\label{SCSN}
S_C &=& \sum_{N_A} p_A(N_A)S[\rho_A^{(N_A)}/\Tr \rho_A^{(N_A)}], \nonumber \\
S_N &=& -\sum_{N_A} p_A(N_A)\,\ln p_A(N_A),
\end{eqnarray}
where $S_C$ is called the configurational entropy and $S_N$ the number entropy. $p_A(N_A)=\Tr \rho_A^{(N_A)}$ denotes the probability to find $N_A$ particles in subsystem $A$ and $\rho_A^{(N_A)}$ is the block of the reduced density matrix $\rho_A$ with $N_A$ particles. We will also consider the R\'enyi number entropy which is defined as
\begin{equation}
\label{Renyi}
S_N^{(\alpha)} = \frac{1}{1 - \alpha} \ln \left\{ \sum_{N_A} \left(p_A(N_A)\right)^{\alpha} \right\}.
\end{equation}
In the limit $\alpha \to 1$, this reduces to the number entropy $S_N$ defined above, while for $\alpha \to 0$ it gives the Hartley number entropy which returns the logarithm of the number of particle-number sectors that the time-evolved wavefunction populates with nonzero weight. Because it weights all populated sectors equally, $S_H=S_{N}^{(\alpha \to 0)}$ is sensitive to configurations which occur with low probability. In other words, the Hartley number entropy captures rare events that may be too small to noticeably affect the number entropy $S_N$, and is a direct probe of how many particles cross the bipartition; it therefore provides a more sensitive measure of how particles spread across the system during the time evolution.

The particle number and Hartley number entropies provide a probe of particle transport; in a truly localized phase, $S_N$ and $S_H$ should saturate rapidly to a value independent of the system size, reflecting no particle transport. However, a series of numerical studies by Kiefer-Emmanouilidis \textit{et al.} \cite{KieferUnanyan2,KieferUnanyan3,KieferUnanyan4,KieferSciPost2022} for the XXZ chain with random magnetic fields demonstrated otherwise. Their findings show that in the putative MBL phase, both the number and the Hartley number entropy continue to grow slowly in time, following a double-logarithmic scaling $S_{N,H} \sim \ln \ln t$ with saturation values that increase logarithmically with system size \cite{KieferUnanyan2,KieferUnanyan3}. Furthermore, the time scale $t_d$ where the double logarithmic scaling breaks down and saturation sets in scales as $t_d \sim e^{L}$ \cite{KieferUnanyan4}, questioning the stability of the MBL phase in the thermodynamic limit in line with concerns previously expressed by Šuntajs \textit{et al.} \cite{SuntajsBonca}. 

This interpretation has been debated however \cite{LuitzBarLev,GhoshZnidaricPRB2022,AceitunoChavezPRL2024,faulend2026}. Ghosh and \v{Z}nidari\v{c}'s study \cite{GhoshZnidaricPRB2022} attributed the slow growth of $S_N$ to rare resonances between localized regions and argued that the resulting dynamics remains compatible with localization. In a Comment, Kiefer-Emmanouilidis \textit{et al.} questioned this interpretation \cite{KieferComment2203}. The resonance model predicts a power-law approach to a system-size independent saturation value, $S_N(t)=\textrm{const}-B/t$. Fits of the numerical data, however, yield a power-law $1/t^b$ with an exponent $b$ that decreases with system size $L$. The data presented in the Comment \cite{KieferComment2203}  are instead described by $S_N\sim\frac{1}{D^3}\ln\ln t$ for all considered system sizes and disorder strengths before finite-size saturation sets in. Furthermore, for a quench from the N\'eel state the saturation values are not constant but rather increase as $\ln L$, both for $S_N$ and for the Hartley number entropy $S_H$. The latter is more sensitive to small fluctuations and is bounded above by $\ln 3$ if only a single particle crosses the cut \cite{KieferUnanyan4,KieferSciPost2022}; the observed saturation values exceed this bound substantially, implying fluctuations involving more than one particle. Finally, for the initial states and strong disorder where the resonance model in Ref.~\cite{GhoshZnidaricPRB2022} describes the data best, the configurational entropy is negligible and $S\approx S_N$, i.e.~the dynamics in the numerically accessible regime is of single-particle type and not related to the question of MBL.

A different perspective was recently put forward in an $\ell$-bit circuit study which demonstrated ultraslow growth of $S_N$ even in models localized by construction, suggesting that the presence of a $S_N\sim\ln\ln t$ growth alone may not be sufficient to rule out an MBL phenomenology \cite{AceitunoChavezPRL2024}. This confirms earlier results in a very similar study for an effective model of interacting Anderson orbitals, for which the particle fluctuations can be shown to be bounded \cite{KieferSciPost2022}. In both of these $\ell$-bit models, $S_N$ and $S_H$ indeed saturate to values that show only a very weak dependence on system size. In particular, the Hartley number entropy saturates for all system sizes to values $\approx 1$ below the bound $\ln 3$ for a single particle crossing the cut (see Suppl.~Mat.~of Ref.~\cite{AceitunoChavezPRL2024}), indicating that multi-particle fluctuations do not occur. This is fundamentally different from the XXZ case with quenched disorder---which the $\ell$-bit results were directly compared with in Ref.~\cite{KieferSciPost2022}---where the saturation values of $S_H$ increase logarithmically with system size and many-particle fluctuations are clearly visible. 

Very recently, another study proposed a microscopic mechanism for small particle fluctuations in a localized phase \cite{faulend2026}. That small particle fluctuations and a corresponding increase of the number entropy do occur even in models localized by construction is consistent with our findings in Ref.~\cite{KieferSciPost2022}. However, the proposed quantum-beat mechanism does not account for the observed multi-particle fluctuations and the logarithmic increase of the saturation value with system size. The predicted differences between quenched disorder and quasi-periodic potentials are also not consistent with the findings presented in the following.

 All three of these studies show the importance of a proper finite-size scaling analysis and of considering the Hartley number entropy instead of concentrating solely on $S_N$. They do not establish that the observed slow particle transport in the XXZ model with quenched disorder can be understood within an $\ell$-bit model or any other localized model. In fact, Ref.~\cite{KieferSciPost2022} reaches the opposite conclusion.

For quasiperiodic potentials, rare regions with anomalously weak disorder do not exist and avalanche instabilities are therefore absent by construction. Based on exact diagonalizations and finite-size extrapolations of the inverse participation ratio, Vu \textit{et al.} \cite{VuHuang} argued that the interacting Aubry-Andr\'e chain is many-body localized for potential strengths $D\gtrsim 4$ and interaction strengths of the order of the hopping. By numerically studying a quasiperiodic chain coupled to a Markovian bath and supported by real-space RG arguments, Tu \textit{et al.}~\cite{TuDasSarma2023} further argued that quasiperiodic systems are more stable against delocalization than randomly disordered ones for $D\gtrsim 16$, potentially up to arbitrarily large system sizes. Recently, however, long-range resonances have been identified also in the quasiperiodic case \cite{PadhanColbois}. Studying the Heisenberg chain in an Aubry-Andr\'e potential, the authors found fat-tailed distributions of longitudinal correlations at long distances and a typical correlation length which grows with system size in an intermediate regime where standard diagnostics indicate MBL, and identified the responsible rare eigenstates as quasi-degenerate pairs of resonant cat states. At even stronger potentials, on the other hand, the distributions become exponential and the correlation length system-size independent, which was interpreted in Ref.~\cite{PadhanColbois} as a standard, stable MBL phase.

In this work, we probe the regime of intermediate to strong disorder by studying the dynamics after a quantum quench in the interacting Aubry-Andr\'e chain. We find that all considered entropies behave qualitatively the same as in the case of quenched disorder: $S(t)\sim\mu\ln t$ and $S_N(t)\sim\nu\ln\ln t$ with growth rates $\mu\sim\nu\sim D^{-\gamma}$ and $\gamma\approx 3$, in close agreement with the value found for random disorder \cite{KieferUnanyan4}. Most importantly, the saturation values of the Hartley number entropy grow with system size as $S_H(t\to\infty)\sim\ln L$ and remain well above the single-particle bound $\ln 3$ even at strong quasiperiodic potentials $D\gtrsim 10$, where avalanches are absent by construction and rare resonances have been argued to become inoperative \cite{PadhanColbois}.

Our paper is organized as follows: In Sec.~\ref{Model}, we introduce the interacting Aubry-Andr\'e chain and discuss the numerical methods used to study its quench dynamics. In Sec.~\ref{Results}, we present our results. The last section discusses these results in comparison to earlier results for quenched disorder and presents our conclusions.

\section{Model and numerical methods}
\label{Model}
We consider spinless fermions on a chain with open boundary conditions of length $L$ with a Hamiltonian given by
\begin{eqnarray}
\label{Ham}
\hat{H} &=&
-J\sum_{j=1}^{L-1} \left(c^\dagger_j c_{j+1} + \mathrm{h.c.}\right) \\
&+& V\sum_{j=1}^{L-1}(n_j-1/2) (n_{j+1}-1/2)  + \sum_{j=1}^{L} D_j (n_j-1/2). \nonumber
\end{eqnarray}
Here $c^{(\dagger)}_j$ is an annihilation (creation) operator of a spinless fermion at site $j$, $n_j=c^\dagger_jc_j$ is the occupation number operator, $J=1$ the hopping amplitude, $V$ the nearest-neighbor interaction, and the Aubry-Andr\'e onsite potential is given by 
\begin{equation}
\label{Aubry}
D_j = D\cos(2\pi\beta j + \phi),
\end{equation}
with amplitude $D$. Here, $\beta$ is chosen to be the golden ratio $(1+\sqrt{5})/2$ which satisfies a Diophantine condition; this ensures that to achieve a near-repetition with a mismatch of order $D\delta$ a distance of order $1/\delta$ is required. In the non-interacting limit $V=0$, the Aubry-Andr\'e model is then mathematically equivalent to the almost Mathieu operator and exhibits a sharp localization transition at $D_c = 2J$ \cite{Jitomirskaya1999}. We work at half filling and bipartition the chain into two equal subsystems $A$ and $B$. We note that the Hamiltonian \eqref{Ham} is equivalent by a Jordan-Wigner transformation to a spin-1/2 XXZ chain with $D_j$ playing the role of site-dependent magnetic fields.

The system is always prepared in a N\'eel state (alternating occupied and unoccupied sites) and then unitarily time-evolved with the Hamiltonian \eqref{Ham}. For system sizes $L \le 18$ we use exact diagonalizations, and for $L=24$ and $L=28$ we use a second-order Suzuki-Trotter decomposition of the time-evolution operator with a time step $\delta t$ chosen such that the discretization error on the time scales we are able to study is negligible compared to the statistical error of averaging over a finite number of potential configurations. We sample the phase offset $\phi$ uniformly in $[0,2\pi]$. The number of realizations of the potential $R$ varies with system size: $R=20000$ for $L=8,10$, $R=8000$ for $L=12,14$, $R=10000$ for $L=16$, $R=798$ for $L=18$, $R=1069$ for $L=24$, and $R=586$ for $L=28$.

To make the time evolution comparable across system sizes, we define for each potential realization the sequence $\{D_j\}$ once on the indices $j=0,1,\ldots,27$ corresponding to the largest system size $L=28$.  For smaller $L$, we truncate symmetrically from both ends so that the chain $L=24$ uses $j=2,\ldots,25$ and so on. The site index $j$ in Eq.~\eqref{Ham} always refers to this absolute index. Therefore, new on-site potentials only appear at the left and right edges of the system when increasing the system size and only influence the late-time dynamics. 

All calculations are performed in double precision, limiting the times that can be explored using exact diagonalizations to $t\lesssim 10^{13}$. The calculations using a Trotter-Suzuki decomposition are limited to even shorter times because of the accumulating discretization errors.

\section{Results}
\label{Results}
\subsection{Disorder dependence}
\label{dis_dep}
\begin{figure}[h!]
    \centering
    \includegraphics[width=0.99\columnwidth]{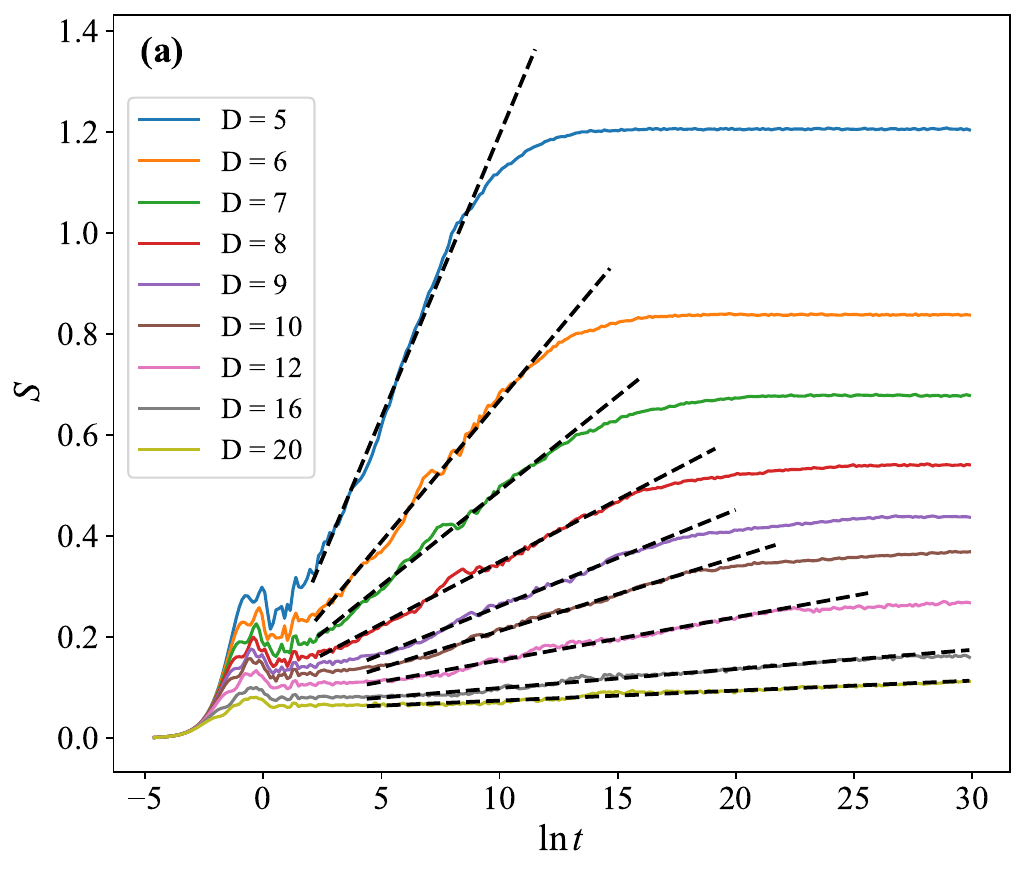}
    \includegraphics[width=0.99\columnwidth]{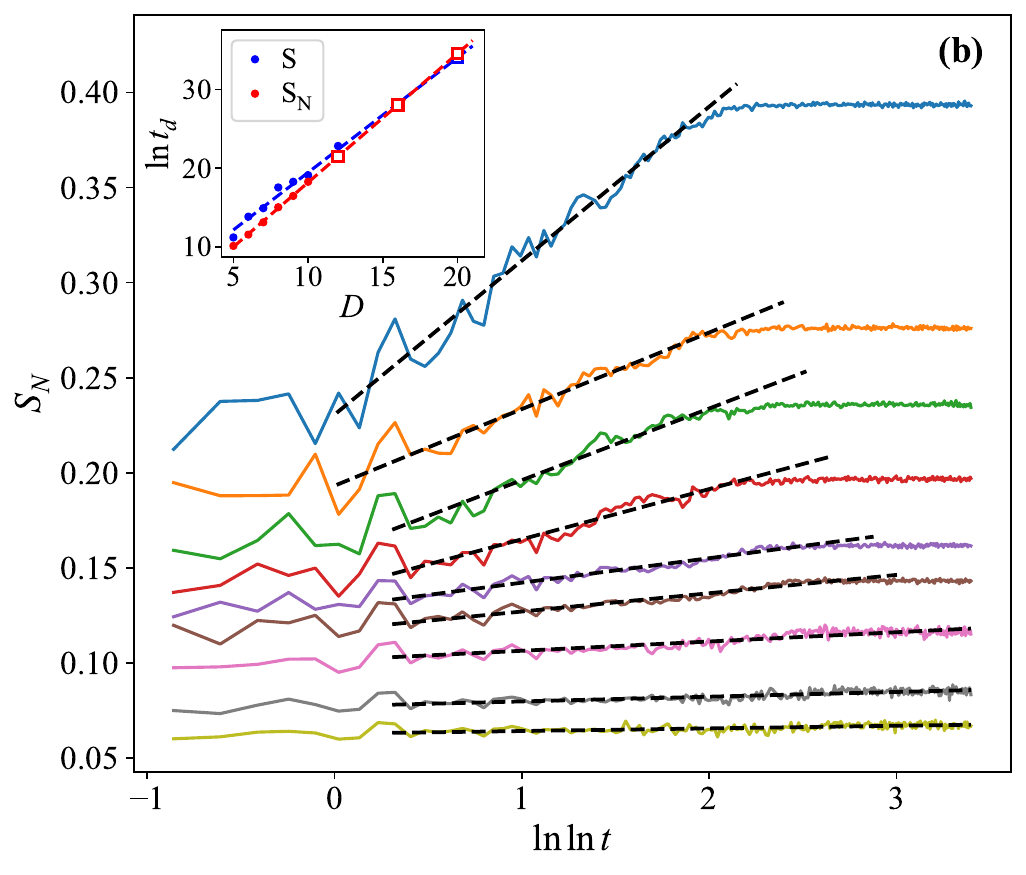}
    \caption{Time evolution of entanglement entropy $S$ and number entropy $S_N$ for system size \(L=16\) and various potential strengths \( D \). The initial state is a N\'eel state and results are averaged over $R=10000$ realizations. (a)  \( S \sim \mu\ln t \) at intermediate times with decreasing $\mu$ for increasing potential strengths \( D \), followed by finite-size saturation at long times. (b) \( S_N \sim \nu\ln \ln t \) at intermediate times followed by saturation. Inset: Deviation times $t_d\sim\text{e}^{aD}$ with almost identical exponents $a$ for $S$ and $S_N$.}
    \label{fig:Sent_Snum_scaling}
\end{figure}

We begin with the time evolution of the entanglement and number entropies in the interacting Aubry-Andr\'e chain for $V=1$ and note that for interaction strengths $V$ of the order of the hopping $J=1$ and potential strengths $D> V$, the results are expected to only weakly depend on the precise value of $V$, see also Sec.~\ref{Int_dep}. Fig.~\ref{fig:Sent_Snum_scaling} shows $S(t)$ and $S_N(t)$ for $L=16$ and potential strengths $D=5$--$20$, averaged over $R=10,000$ realizations. The entanglement entropy shows three distinct regimes in time. An initial fast increase at times of order of the hopping $J$, followed by a slower increase consistent with $S(t)\sim\mu\ln t$, see the fits (dashed lines) in Fig.~\ref{fig:Sent_Snum_scaling}(a). This is followed by a saturation. The prefactor $\mu$ of the logarithmic growth monotonically decreases with increasing potential strength $D$, as does the saturation value. The growth of the number entropy $S_N(t)$ also shows three regimes. At short times, there is overall growth but with strong fluctuations, followed by a regime where $S_N\sim\nu\ln\ln t$ holds, and finally, again, a regime where $S_N$ saturates. As for the entanglement entropy, the prefactor $\nu$ of the double logarithmic growth as well as the saturation values are monotonically decreasing functions of the potential strength $D$. 

The inset of Fig.~\ref{fig:Sent_Snum_scaling}(b) shows the deviation times $t_d$ for both $S$ and $S_N$ as a function of $D$. Here, the deviation time is defined as the time at which the data deviates by a fixed threshold of 10\% from the logarithmic and double-logarithmic fits, respectively. We note that while the exact values of $t_d$ do depend on the threshold chosen, the scaling as a function of $D$ is generic. For both entropies, the deviation times increase as $\ln t_d\sim a D$ with the potential strength $D$ and with nearly identical exponents $a$. There is, in particular, no indication that $t_d$ continues to grow for the entanglement entropy while it starts to saturate for the number entropy, a behavior which might be expected if we cross from an ergodic into an MBL phase at some critical potential strength $D_c^{\textit{MBL}}$, where particle spreading stops but dephasing continues. The data are consistent with the relation $S_N \sim \ln S$ established for the quenched disorder case \cite{KieferSciPost2020,KieferUnanyan2,KieferUnanyan3,KieferUnanyan4}, and we find no qualitative difference between the quasiperiodic and the previously studied quenched disorder case when considering the relation between entanglement and number entropy as well as the deviation times. 
\begin{figure}[h!]
    \centering
    \includegraphics[width=0.99\columnwidth]{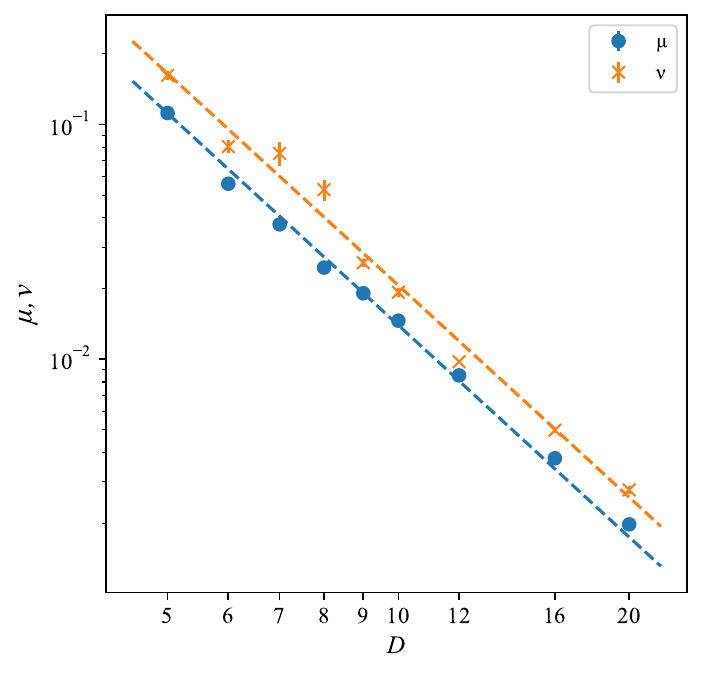}
    \caption{Scaling of the parameters $\mu$ and $\nu$ as a function of the potential strength $D$ extracted from fits $S\sim\mu\ln t$ and $S_N\sim\nu\ln\ln t$. Both parameters show the same power-law scaling $\mu\sim\nu\sim D^{-\gamma}$ consistent with $\gamma\approx 3$, see dashed lines.}
    \label{fig:plot_u_v}
\end{figure}
These results are further supported by a quantitative analysis of the logarithmic and double logarithmic fits, respectively. As shown in Fig.~\ref{fig:plot_u_v}, both prefactors $\mu$ and $\nu$ decrease as a power law, $\mu,\nu \sim D^{-\gamma}$, with increasing potential strength $D$. The exponent $\gamma$ is consistent with $\gamma=3$ (see dashed lines in Fig.~\ref{fig:plot_u_v}) and also consistent with the exponent found in the quenched disorder case studied in Ref.~\cite{KieferUnanyan3}. As for every numerical study, we cannot exclude that the relation $S_N\sim \ln S$ breaks down for even larger strengths of the potential $D$ but for the time scales, lengths,  and potential strengths we can study we find no qualitative differences in the behavior of the entanglement and number entropy for the quasi-periodic Aubry-Andr\'e chain as compared to the XXZ chain with quenched disorder.
\begin{figure*}[t]
    \centering
    \includegraphics[width=0.85\textwidth]{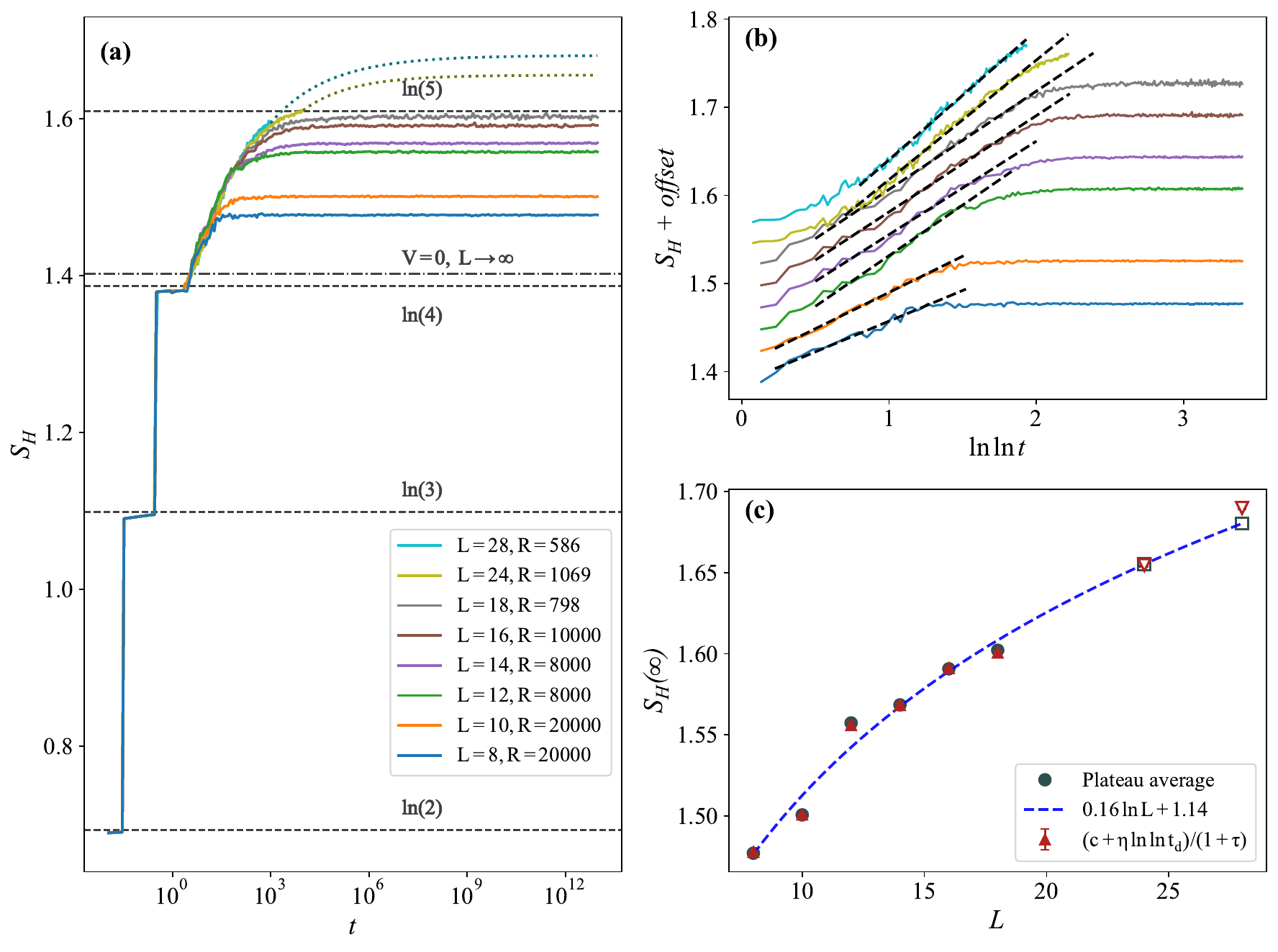}
    \caption{(a) $S_H(t)$ for $D=7$ and system sizes $L=8, 10, 12, 14, 16, 18$ obtained from ED and $L=24, 28$ from a Trotter-Suzuki decomposition of the time evolution operator. The continuations of the $L=24, 28$ curves are guides to the eye connecting the data to the saturation values predicted in panel (c). The number of potential realizations $R$ is given in the legend. The non-interacting model ($V=0$) saturates at $S_H\approx 1.40$ for $L\to\infty$. (b) The same data shown against $\ln \ln t$ with a constant offset $0.025$ and fits to $S_H(t) = c + \eta \ln \ln t$; the fit parameters $c$, $\eta$ are extracted independently for each $L$. (c) Saturation values $S_H(t\to\infty)$ as a function of $L$. Filled blue circles: late-time average of the $L=8\text{--}18$ data. Open blue squares: extrapolation of the $0.16 \ln L + 1.14$ fit to $L=24, 28$. Filled red triangles: Self-consistent estimate $(c + \eta \ln\ln t_d)/(1+\tau)$, where $t_d$ is the time at which the data first deviate from the fit by $\tau\, S_H(t\to\infty)$ with $\tau = 0.03$. Open red triangles: same estimate for $L=24, 28$, with $t_d$ obtained from the scaling $\ln t_d = a_L L + b_L$ established for $L=8$ to $18$.}
    \label{fig:SHart_V0}
\end{figure*}

As we have already alluded to in the introduction it is, however, extremely important to also consider measures that are more sensitive to small particle number fluctuations than the number entropy and to perform an analysis of how the results depend on system size. To address both of these points, we consider next the Hartley number entropy $S_H$ which is formally defined as the limit of the R\'enyi number entropy \eqref{Renyi} for $\alpha\to 0$. Numerically, however, taking this limit exactly is not possible. Specifically, for $\alpha\to 0$ every particle number sector would fully be counted for arbitrarily small occupation even if this occupation is smaller than double precision. Numerically, we therefore fix $\alpha=10^{-3}$ instead of trying to take the limit $\alpha\to 0$. This will only result in a slight overall suppression of the Hartley number entropy but will not change the results qualitatively. Second, we introduce a cutoff where we only consider sectors with $p_A(N_A)>10^{-10}$ and we then normalize the sum of the probabilities which fulfill this condition to $1$. Choosing different cutoffs will mostly correspond to a renormalization of time: for a smaller cutoff a sector starts to contribute earlier while for a larger cutoff it starts to contribute later. As a consequence, also the saturation value will shift by a constant offset that depends on the chosen cutoff. However, the qualitative structure of the Hartley number entropy, and in particular the scaling of the saturation value with system size, remains unchanged. We call the numerical Hartley number entropy obtained using this procedure $S_H(t)$. 

Fig.~\ref{fig:SHart_V0} shows $S_H(t)$ at $D = 7$, well above the non-interacting localization threshold $D_c=2$, for system sizes $L=8$ to $L=28$. The Hartley number entropy as defined above essentially gives the logarithm of the number of the particle sectors with $p_A(N_A)$ above the cutoff. As a result of the cutoff, the time evolution in panel~(a) shows a clear staircase structure, with the data passing through $\ln 2$, $\ln 3$, and $\ln 4$ as additional particle-number sectors become populated above the threshold. The maximum value of $S_H$ in a half-filled system which is cut in half is $\ln(L/2 + 1)$. We find again three dynamical regimes. First, $S_H$ increases rapidly and even for the smallest system size investigated, $L=8$, quickly reaches values larger than $\ln 3$. At this point, particle number fluctuations include sectors where two particles have crossed the cut. $S_H$ also quickly crosses $S_H\approx 1.40$ which is the saturation value in the thermodynamic limit for the non-interacting model with $D=7$. Beyond this initial regime, the data enter a slow growth phase in which $S_H(t)\sim\ln\ln t$, i.e.~the same qualitative increase as the number entropy $S_N(t)$, see Fig.~\ref{fig:SHart_V0}(b). We can fit the growth in this regime by
\begin{equation}
S_H(t) = c + \eta \ln\ln t,
\label{eq:Hartley_lnlnt_fit}
\end{equation}
with $c, \eta$ extracted independently for each system size $L$. Finally, $S_H$ starts to saturate to a length-dependent value.

In Fig.~\ref{fig:SHart_V0}(c), the saturation of $S_H$ as a function of length $L$ is further analyzed. For $L = 8-18$, where the late time plateau is directly accessible (filled blue circles), the saturation values increase with $L$ and are well described by a logarithmic growth with system size (see dashed line in Fig.~\ref{fig:SHart_V0}(c)). For $L = 24$ and $L = 28$, the Trotter-Suzuki  dynamics has not reached the saturation plateau within the accessible time window, and we estimate the saturation value for these two lengths by two complementary methods. The first is the extrapolation of the logarithmic fit established for the smaller systems (open blue squares). The second is a self-consistency estimate, $(c + \eta \ln\ln t_d)/(1+\tau)$, where $t_d$ is the time at which the data first deviate from the $\ln\ln t$ fit in the intermediate time regime by a fraction $\tau$ of the saturation value (filled red triangles for $L \le 18$, open red triangles for $L = 24, 28$, with $t_d$ for the largest two systems obtained from the scaling $\ln t_d = a_L L + b_L$ established for $L = 8$ to $18$). The two estimates agree within better than $1\%$ for all system sizes $L$, lending confidence to the extrapolated values at $L = 24, 28$.

In a many-body localized phase, the model is by definition unitarily equivalent to an effective model of quasi-local conserved charges. As a result, particle fluctuations across a cut should be limited and measures such as the number and Hartley number entropies ultimately have to converge to system-size independent values. For the XXZ model with quenched disorder such an effective model was constructed in \cite{KieferSciPost2022} and it was shown that in such an effective model a saturation of the Hartley number entropy independent of system size indeed occurs for sufficiently large systems. In contrast, the full XXZ model with quenched disorder showed the same $\ln L$ dependence of the saturation values we observe here in the quasi-periodic Aubry-Andr\'e chain; the latter is shown in Fig.~\ref{fig:SHart_V0}(c). We therefore conclude that for the studied potential strengths and system sizes the results for the Aubry-Andr\'e chain are qualitatively similar to the XXZ chain with quenched disorder and, in particular, inconsistent with a stable MBL phase at the considered strength of the disorder potential of $D=7$.

\begin{figure}[h]
    \centering
    \includegraphics[width=0.99\columnwidth]{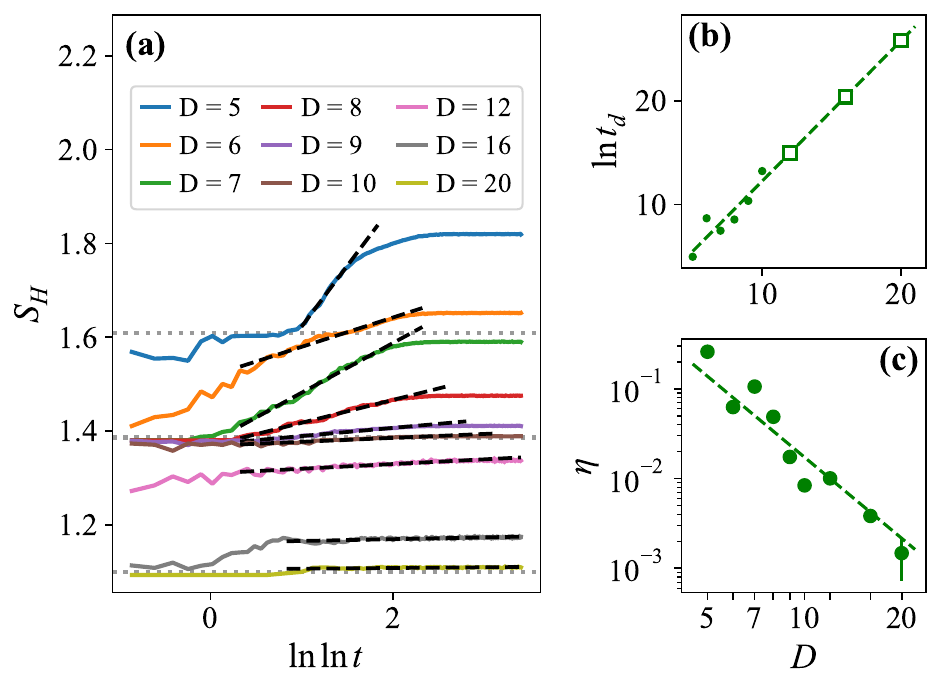}
    \caption{Hartley number entropy $S_H(t)$ of the Aubry-Andr\'e chain with $V=1$ and $L=16$ for disorder strength $D=5-20$, averaged over 10,000 disorder realizations. (a) $S_H\sim\eta\ln\ln t$ at intermediate times with decreasing $\eta$ for increasing potential strength. (b) Deviation times $t_d$, defined as times at which the data deviate from the $\ln\ln$-fit by $10\%$, scale as $t_d\sim \mathrm{e}^{aD}$. For $D=12,16,20$ the simulations do not reach saturation and the values are obtained by extrapolation. (c) The prefactor $\eta$ as a function of $D$. The dashed line indicates $\eta\sim D^{-3}$.}
    \label{fig:plot_Hartley_D9_D10}
\end{figure}

As a next step, we therefore also consider $S_H$ for even larger values of $D$ where other numerical studies using different measures have argued in favour of a stable MBL phase \cite{TuDasSarma2023,PadhanColbois}. Note that in the convention of Ref.~\cite{PadhanColbois} the potential strength is given by $h=D/2$. Fig.~\ref{fig:plot_Hartley_D9_D10}(a) shows $S_H$ for $D\in[5,20]$ and system size $L=16$. We observe that the $S_H\sim\eta\ln\ln t$ scaling persists for all $D$; only the prefactor changes as $\eta\sim D^{-3}$, see Fig.~\ref{fig:plot_Hartley_D9_D10}(c). This scaling is the same as the one we observed for the entanglement and number entropies, see Fig.~\ref{fig:plot_u_v}. Also consistent with the previous results is that the deviation time increases exponentially $t_d\sim \text{e}^{aD}$, see Fig.~\ref{fig:plot_Hartley_D9_D10}(b). We note, furthermore, that we observe the same logarithmic scaling of the saturation values with length $L$ as shown in Fig.~\ref{fig:SHart_V0}(c) for potentials $D>7$ where we, however, can only consider system sizes $L\leq 16$ because of the saturation times increasing exponentially with $D$ (data not shown). We conclude that we do not see any qualitative changes in the behavior of $S_H$ for $D\geq 12$, corresponding to $h\geq 6$, where Ref.~\cite{PadhanColbois} finds exponentially suppressed tails of the long-distance correlations and a system-size independent correlation length, i.e.~no signatures of long-range resonances. 

\subsection{Interaction dependence}
\label{Int_dep}
\begin{figure}[h]
    \centering
    \includegraphics[width=0.50\textwidth]{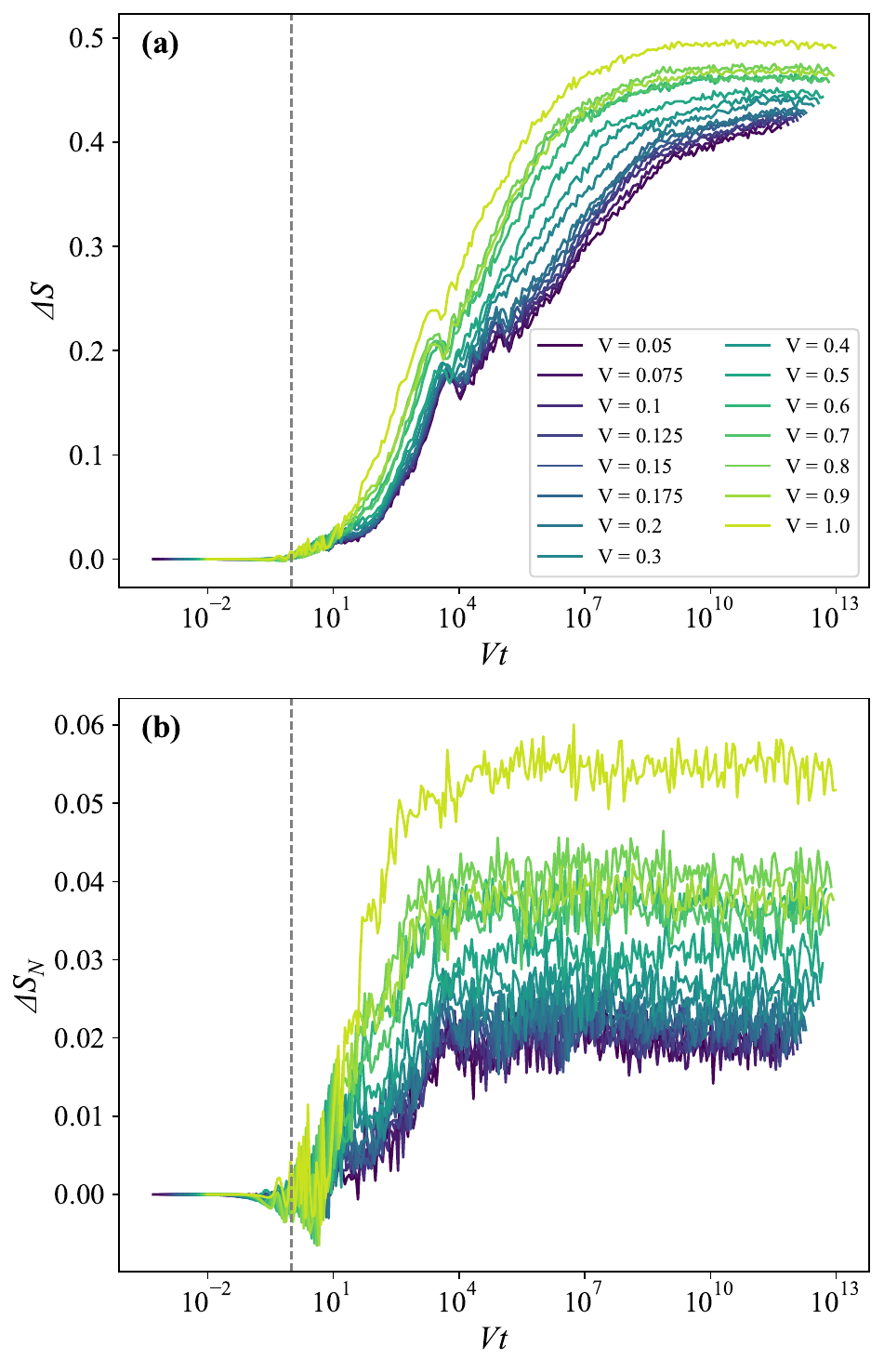}
    \caption{(a) Entanglement entropy $\Delta S= S-S_0$, where $S_0$ is the entanglement entropy for $V=0$, plotted as a function of $Vt$ for interaction strengths $V\in[0.05,1]$ and $D=7$. (b) Same as in (a) for the number entropy $\Delta S_N$.}
\label{fig:V1}
\end{figure}

In addition to the disorder dependence studied in the previous subsection, it is also useful to study the interaction dependence. This will highlight the differences between the Anderson and the many-body case and the mechanisms responsible for the logarithmic growth of the entanglement entropy and the double logarithmic growths of the number and Hartley number entropies, respectively. The entanglement entropy $\Delta S = S-S_0$, where $S_0$ is the entanglement entropy of the non-interacting model ($V=0$), is shown in Fig.~\ref{fig:V1}(a) as a function of the scaled time $Vt$. Two conclusions can be drawn from the numerical data: (i) $\Delta S$ starts to deviate from zero at times $Vt^*\sim 1$, i.e., the onset time $t^*$ where many-body effects become visible in the entanglement entropy scales as $t^*\sim 1/V$. (ii) The slope of the logarithmic increase, $S\sim\mu\ln t$, shows only a weak dependence on the interaction strength $V$. 

\begin{figure}[h]
    \centering
    \includegraphics[width=0.50\textwidth]{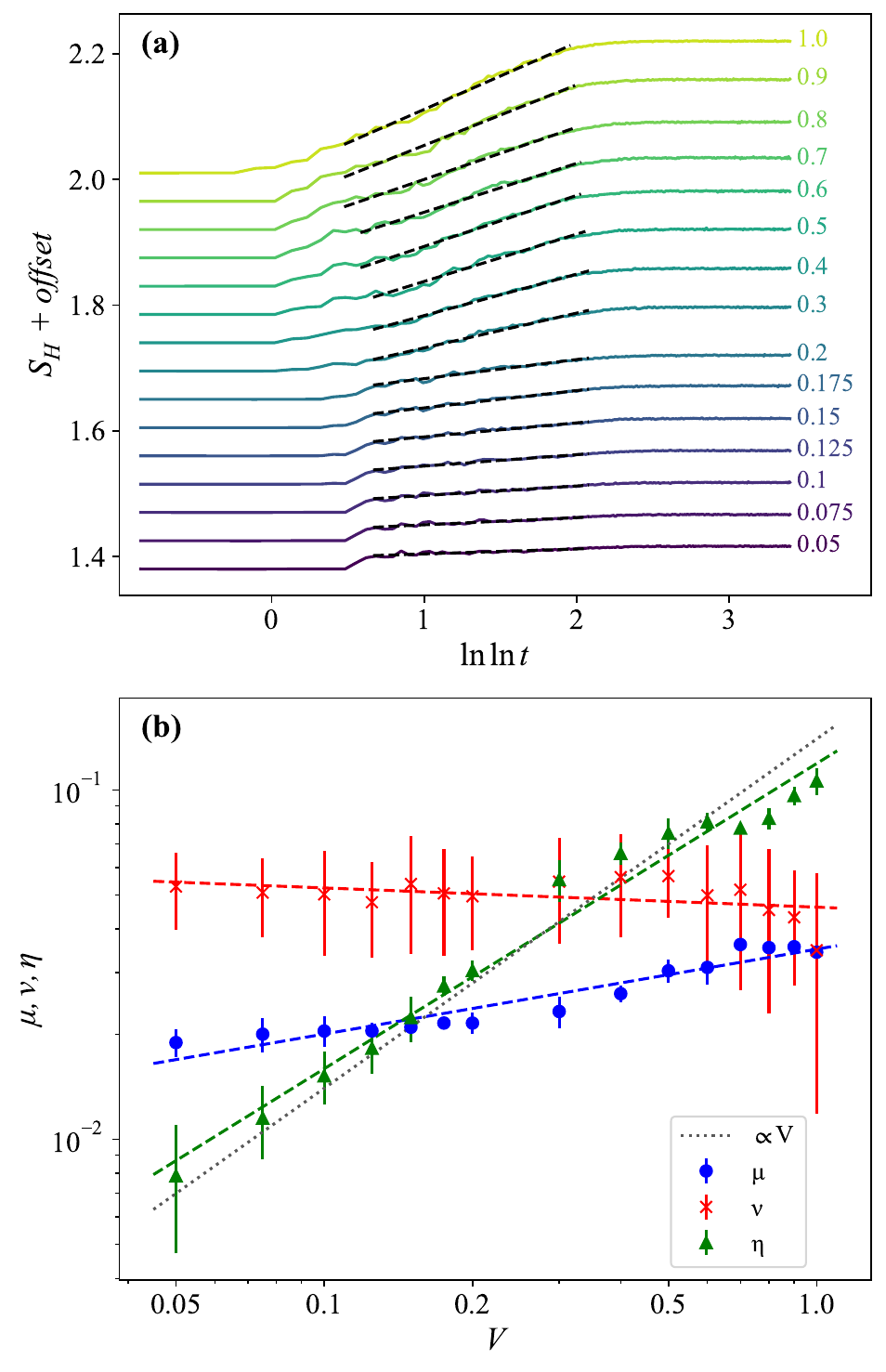}
    \caption{(a) $S_H$ for interaction strengths $V\in [0.05,1]$ and $D=7$. Note that each curve has a vertical offset of $0.045$. The dashed lines are $\ln\ln t$ fits at intermediate times. (b) Prefactors of the logarithmic (double logarithmic) growth of $S$ ($S_N,S_H$) versus interaction strength $V$. The dashed lines are power-law fits, the dotted line is a linear guide to the eye.}
\label{fig:V2}
\end{figure}

Both of these points can be understood in the standard dephasing picture \cite{AbaninRev2019}. Sites a distance $r$ apart feel an effective interaction $J_{\textrm{eff}}\sim V\text{e}^{-r/\xi}$ where $\xi$ is some effective localization length. For nearby sites, $r\ll\xi$, this leads to an onset of dephasing at times $t^*\sim J_{\textrm{eff}}^{-1}\sim 1/V$ consistent with the numerical data. The length scale, over which dephasing is active, grows as $r(t)\sim\xi\ln(Vt)$. For the entanglement entropy it follows that
\begin{equation}
    \label{S_deph}
    S(t)\sim r(t)\sim \xi(V,D)\ln(t) \, .
\end{equation}
This predicts that the slope of the logarithmic increase only depends on $V$ indirectly through the effective localization length $\xi(V,D)$. The latter is expected to change strongly with disorder but to show only a weak dependence on $V$. This is again consistent with the numerical data, see also Fig.~\ref{fig:V2}(b). 

What is surprising and not consistent with this naive picture of a pure dephasing between frozen particles is the increase in the number entropy, see Fig.~\ref{fig:V1}(b).
As for the entanglement entropy, we find again that the particle dynamics starts to deviate from the non-interacting case at times $t^*\sim 1/V$ while the double logarithmic increase at intermediate times shows only a weak dependence on interaction strength $V$. This suggests that the dynamics setting in at times $t^*$ is not a pure dephasing dynamics but goes hand in hand with an increase in particle fluctuations between the two subsystems. A separation between entanglement and particle dynamics, as would be expected in a pure dephasing picture deep in an MBL phase, is not observed consistent with the investigation of the disorder dependence in the previous subsection. 

Finally, we show results for the interaction dependence of the Hartley number entropy $S_H$ in Fig.~\ref{fig:V2}(a). We find for all interaction strengths an intermediate time interval with $S_H\sim\eta\ln\ln t$. Here, the onset of the scaling shows only a weak dependence on $V$ while the prefactor, $\eta\sim V$, changes continuously, see Fig.~\ref{fig:V2}(b). This can be explained as follows: while $S_N$ is only sensitive to major changes of the particle number distribution, $S_H$ in contrast is sensitive to the tails of the distribution. The results show that additional particle sectors start contributing to $S_H$ already at times $t\sim 1/J$ while the speed at which new sectors get filled above the threshold $p_A(N_A)=10^{-10}$ is set by $V$. 

\section{Conclusions}
\label{Conclusions}
There are two known mechanisms that can destabilize a potential MBL phase. On the one hand, there can be energy resonances and regions that are almost thermal, leading to avalanche instabilities. This mechanism is energy-based and affects the denominators in any scheme that tries to transform the microscopic Hamiltonian to an effective Hamiltonian based on quasi-local charges. On the other hand, there is a path-entropy obstruction caused by the exponentially large size of the operator space in an interacting many-body system that leads to extensive branching when changing the operator basis. This mechanism affects the numerators of any transformation scheme and is independent of the resonance and avalanche mechanism.

If resonances and avalanches are the dominant mechanism destabilizing MBL phases, then one would expect to see qualitative differences in the dynamics of a system with quenched disorder as compared to a system with a quasi-periodic potential, where avalanches are absent by construction and the statistics of resonances is strongly modified. Dynamical indicators which should be able to see such changes are the growth of the particle number and the Hartley number entropies after a quench. For the XXZ chain with quenched disorder we have established previously that these indicators point to a continuing slow delocalization of particles even at very strong disorder \cite{KieferUnanyan2,KieferUnanyan3,KieferUnanyan4,KieferComment2203}.

Here we have contrasted these results with a study of the same indicators for the interacting Aubry-Andr\'e chain, a model with a quasi-periodic potential. We have chosen the frequency of the potential to be the golden ratio such that in order to find near-repetitions of the potential with a mismatch as small as $\mathcal{O}(D\delta)$ one has to consider distances of $\mathcal{O}(1/\delta)$, i.e., this is the badly approximable case in which approximate repetitions are as poor as possible for any irrational frequency. Furthermore, for each realization of the potential---each realization consists of picking the potential with a specific phase factor $\phi\in [0,2\pi]$---we have defined the sequence $\{D_j\}$ once on the sites of the largest system size considered and truncated it symmetrically for smaller $L$, so that the potential landscape around the cut is identical for all system sizes.

We have found that in the intermediate time regime---following a fast initial growth of the entropy and before finite-size saturation sets in---the entanglement entropy is well described by $S\sim\mu\ln t$ and the number entropy by $S_N\sim\nu\ln\ln t$ for all potential strengths $D$ and system sizes $L$ we considered. Both prefactors scale as $\mu,\nu\sim D^{-\gamma}$ with $\gamma\approx 3$ and thus remain tied to each other. In particular, there is no indication of the particle fluctuations saturating while the entanglement entropy keeps growing, as expected for an MBL phase. This behavior, including the power-law scaling  of the prefactors $\sim D^{-3}$, is the same as for the quenched disorder case. 

To be more sensitive to small particle number fluctuations across the cut, we also considered the Hartley number entropy $S_H$. This quantity significantly exceeds at long times the bound $\ln 3$ and also the non-interacting value, ruling out simple single-particle resonances as an explanation. Furthermore, we find that the saturation values increase with system size as $\ln L$ for all potential strengths we investigated. In particular, we do not see a transition into a phase where the saturation value becomes independent of system size even for potential strengths consistent with what other recent numerical studies have identified as the critical potential strength for the MBL transition. Notably, the $\ln L$ growth persists at $D\geq 12$, where Ref.~\cite{PadhanColbois} finds no signatures of long-range resonances in the eigenstate correlations. Since in a localized-by-construction $\ell$-bit model the saturation values of $S_N$ and $S_H$ become independent of system size \cite{KieferSciPost2022,AceitunoChavezPRL2024}, the observed $\ln L$ growth is inconsistent with a stable MBL phase. We also note that at intermediate times $S_H\sim\eta\ln\ln t$ holds with $\eta\sim D^{-3}$, i.e., the exact same scaling as a function of disorder as observed for $S$ and $S_N$.

We also considered the interaction dependence of these three entropies at fixed disorder strength. The results for $S$ are consistent with the standard dephasing picture with an onset time of the logarithmic growth scaling as $t^*\sim 1/V$ and the slope of the logarithmic increase being only weakly dependent on $V$. However, we find again, and in contrast to a simple dephasing picture, that the number entropy closely follows the behavior of the entanglement entropy, $S_N\sim\ln S$, for all $V$. I.e., $t^*$ not only indicates the onset of dephasing due to the interaction but also the onset of substantial particle number fluctuations. 

The most important conclusion of our study is that the quenched disorder and the quasi-periodic case show very similar behavior---including an even quantitative agreement in the scaling of the logarithmic and double logarithmic growth of the entanglement and number entropies with potential strength and of the deviation times---despite the fundamentally different potential spectra of the two models. Moreover, the slow particle transport persists in a regime where avalanches are excluded by construction and where dedicated eigenstate diagnostics detect no long-range resonances \cite{PadhanColbois}, i.e., where both known energy-based mechanisms appear to be inoperative. We interpret our results as pointing towards a universal obstruction against the stability of an MBL phase in the thermodynamic limit that is independent of resonances and avalanches. Such an obstruction was recently discussed in Refs.~\cite{WeisseGerstnerSirker,SirkerObstruction} and springs from the generic extensive branching of operator content in many-body systems. 

%\bibliographystyle{apsrev4-2}
%\bibliography{bib,Literatur}

\begin{thebibliography}{41}%
\makeatletter
\providecommand \@ifxundefined [1]{%
 \@ifx{#1\undefined}
}%
\providecommand \@ifnum [1]{%
 \ifnum #1\expandafter \@firstoftwo
 \else \expandafter \@secondoftwo
 \fi
}%
\providecommand \@ifx [1]{%
 \ifx #1\expandafter \@firstoftwo
 \else \expandafter \@secondoftwo
 \fi
}%
\providecommand \natexlab [1]{#1}%
\providecommand \enquote  [1]{``#1''}%
\providecommand \bibnamefont  [1]{#1}%
\providecommand \bibfnamefont [1]{#1}%
\providecommand \citenamefont [1]{#1}%
\providecommand \href@noop [0]{\@secondoftwo}%
\providecommand \href [0]{\begingroup \@sanitize@url \@href}%
\providecommand \@href[1]{\@@startlink{#1}\@@href}%
\providecommand \@@href[1]{\endgroup#1\@@endlink}%
\providecommand \@sanitize@url [0]{\catcode `\\12\catcode `\$12\catcode
  `\&12\catcode `\#12\catcode `\^12\catcode `\_12\catcode `\%12\relax}%
\providecommand \@@startlink[1]{}%
\providecommand \@@endlink[0]{}%
\providecommand \url  [0]{\begingroup\@sanitize@url \@url }%
\providecommand \@url [1]{\endgroup\@href {#1}{\urlprefix }}%
\providecommand \urlprefix  [0]{URL }%
\providecommand \Eprint [0]{\href }%
\providecommand \doibase [0]{https://doi.org/}%
\providecommand \selectlanguage [0]{\@gobble}%
\providecommand \bibinfo  [0]{\@secondoftwo}%
\providecommand \bibfield  [0]{\@secondoftwo}%
\providecommand \translation [1]{[#1]}%
\providecommand \BibitemOpen [0]{}%
\providecommand \bibitemStop [0]{}%
\providecommand \bibitemNoStop [0]{.\EOS\space}%
\providecommand \EOS [0]{\spacefactor3000\relax}%
\providecommand \BibitemShut  [1]{\csname bibitem#1\endcsname}%
\let\auto@bib@innerbib\@empty
%</preamble>
\bibitem [{\citenamefont {Anderson}(1958)}]{Anderson58}%
  \BibitemOpen
  \bibfield  {author} {\bibinfo {author} {\bibfnamefont {P.~W.}\ \bibnamefont
  {Anderson}},\ }\bibfield  {title} {\bibinfo {title} {Absence of diffusion in
  certain random lattices},\ }\href {https://doi.org/10.1103/PhysRev.109.1492}
  {\bibfield  {journal} {\bibinfo  {journal} {Phys. Rev.}\ }\textbf {\bibinfo
  {volume} {109}},\ \bibinfo {pages} {1492} (\bibinfo {year}
  {1958})}\BibitemShut {NoStop}%
\bibitem [{\citenamefont {Basko}\ \emph {et~al.}(2006)\citenamefont {Basko},
  \citenamefont {Aleiner},\ and\ \citenamefont {Altshuler}}]{AleinerAltshuler}%
  \BibitemOpen
  \bibfield  {author} {\bibinfo {author} {\bibfnamefont {D.}~\bibnamefont
  {Basko}}, \bibinfo {author} {\bibfnamefont {I.}~\bibnamefont {Aleiner}},\
  and\ \bibinfo {author} {\bibfnamefont {B.}~\bibnamefont {Altshuler}},\
  }\bibfield  {title} {\bibinfo {title} {Metal-insulator transition in a weakly
  interacting many-electron system with localized single-particle states},\
  }\href {https://doi.org/http://dx.doi.org/10.1016/j.aop.2005.11.014}
  {\bibfield  {journal} {\bibinfo  {journal} {Ann. Phys.}\ }\textbf {\bibinfo
  {volume} {321}},\ \bibinfo {pages} {1126} (\bibinfo {year}
  {2006})}\BibitemShut {NoStop}%
\bibitem [{\citenamefont {Nandkishore}\ and\ \citenamefont
  {Huse}(2015)}]{NandkishoreHuse}%
  \BibitemOpen
  \bibfield  {author} {\bibinfo {author} {\bibfnamefont {R.}~\bibnamefont
  {Nandkishore}}\ and\ \bibinfo {author} {\bibfnamefont {D.~A.}\ \bibnamefont
  {Huse}},\ }\bibfield  {title} {\bibinfo {title} {Many-body localization and
  thermalization in quantum statistical mechanics},\ }\href
  {https://doi.org/10.1146/annurev-conmatphys-031214-014726} {\bibfield
  {journal} {\bibinfo  {journal} {Annual Review of Condensed Matter Physics}\
  }\textbf {\bibinfo {volume} {6}},\ \bibinfo {pages} {15} (\bibinfo {year}
  {2015})}\BibitemShut {NoStop}%
\bibitem [{\citenamefont {Abanin}\ \emph {et~al.}(2019)\citenamefont {Abanin},
  \citenamefont {Altman}, \citenamefont {Bloch},\ and\ \citenamefont
  {Serbyn}}]{AbaninRev2019}%
  \BibitemOpen
  \bibfield  {author} {\bibinfo {author} {\bibfnamefont {D.~A.}\ \bibnamefont
  {Abanin}}, \bibinfo {author} {\bibfnamefont {E.}~\bibnamefont {Altman}},
  \bibinfo {author} {\bibfnamefont {I.}~\bibnamefont {Bloch}},\ and\ \bibinfo
  {author} {\bibfnamefont {M.}~\bibnamefont {Serbyn}},\ }\bibfield  {title}
  {\bibinfo {title} {Colloquium: Many-body localization, thermalization, and
  entanglement},\ }\href {https://doi.org/10.1103/RevModPhys.91.021001}
  {\bibfield  {journal} {\bibinfo  {journal} {Rev. Mod. Phys.}\ }\textbf
  {\bibinfo {volume} {91}},\ \bibinfo {pages} {021001} (\bibinfo {year}
  {2019})}\BibitemShut {NoStop}%
\bibitem [{\citenamefont {\ifmmode \check{Z}\else
  \v{Z}\fi{}nidari\ifmmode~\check{c}\else \v{c}\fi{}}\ \emph
  {et~al.}(2008)\citenamefont {\ifmmode \check{Z}\else
  \v{Z}\fi{}nidari\ifmmode~\check{c}\else \v{c}\fi{}}, \citenamefont {Prosen},\
  and\ \citenamefont {Prelov\ifmmode~\check{s}\else
  \v{s}\fi{}ek}}]{ZnidaricProsen}%
  \BibitemOpen
  \bibfield  {author} {\bibinfo {author} {\bibfnamefont {M.}~\bibnamefont
  {\ifmmode \check{Z}\else \v{Z}\fi{}nidari\ifmmode~\check{c}\else
  \v{c}\fi{}}}, \bibinfo {author} {\bibfnamefont {T.}~\bibnamefont {Prosen}},\
  and\ \bibinfo {author} {\bibfnamefont {P.}~\bibnamefont
  {Prelov\ifmmode~\check{s}\else \v{s}\fi{}ek}},\ }\bibfield  {title} {\bibinfo
  {title} {Many-body localization in the heisenberg $xxz$ magnet in a random
  field},\ }\href {https://doi.org/10.1103/PhysRevB.77.064426} {\bibfield
  {journal} {\bibinfo  {journal} {Phys. Rev. B}\ }\textbf {\bibinfo {volume}
  {77}},\ \bibinfo {pages} {064426} (\bibinfo {year} {2008})}\BibitemShut
  {NoStop}%
\bibitem [{\citenamefont {Pal}\ and\ \citenamefont {Huse}(2010)}]{PalHuse}%
  \BibitemOpen
  \bibfield  {author} {\bibinfo {author} {\bibfnamefont {A.}~\bibnamefont
  {Pal}}\ and\ \bibinfo {author} {\bibfnamefont {D.~A.}\ \bibnamefont {Huse}},\
  }\bibfield  {title} {\bibinfo {title} {Many-body localization phase
  transition},\ }\href {https://doi.org/10.1103/PhysRevB.82.174411} {\bibfield
  {journal} {\bibinfo  {journal} {Phys. Rev. B}\ }\textbf {\bibinfo {volume}
  {82}},\ \bibinfo {pages} {174411} (\bibinfo {year} {2010})}\BibitemShut
  {NoStop}%
\bibitem [{\citenamefont {Oganesyan}\ and\ \citenamefont
  {Huse}(2007)}]{OganesyanHuse}%
  \BibitemOpen
  \bibfield  {author} {\bibinfo {author} {\bibfnamefont {V.}~\bibnamefont
  {Oganesyan}}\ and\ \bibinfo {author} {\bibfnamefont {D.~A.}\ \bibnamefont
  {Huse}},\ }\bibfield  {title} {\bibinfo {title} {Localization of interacting
  fermions at high temperature},\ }\href
  {https://doi.org/10.1103/PhysRevB.75.155111} {\bibfield  {journal} {\bibinfo
  {journal} {Phys. Rev. B}\ }\textbf {\bibinfo {volume} {75}},\ \bibinfo
  {pages} {155111} (\bibinfo {year} {2007})}\BibitemShut {NoStop}%
\bibitem [{\citenamefont {Luitz}\ \emph {et~al.}(2015)\citenamefont {Luitz},
  \citenamefont {Laflorencie},\ and\ \citenamefont {Alet}}]{Luitz1}%
  \BibitemOpen
  \bibfield  {author} {\bibinfo {author} {\bibfnamefont {D.~J.}\ \bibnamefont
  {Luitz}}, \bibinfo {author} {\bibfnamefont {N.}~\bibnamefont {Laflorencie}},\
  and\ \bibinfo {author} {\bibfnamefont {F.}~\bibnamefont {Alet}},\ }\bibfield
  {title} {\bibinfo {title} {Many-body localization edge in the random-field
  heisenberg chain},\ }\href {https://doi.org/10.1103/PhysRevB.91.081103}
  {\bibfield  {journal} {\bibinfo  {journal} {Phys. Rev. B}\ }\textbf {\bibinfo
  {volume} {91}},\ \bibinfo {pages} {081103} (\bibinfo {year}
  {2015})}\BibitemShut {NoStop}%
\bibitem [{\citenamefont {Luitz}\ \emph {et~al.}(2016)\citenamefont {Luitz},
  \citenamefont {Laflorencie},\ and\ \citenamefont {Alet}}]{Luitz2}%
  \BibitemOpen
  \bibfield  {author} {\bibinfo {author} {\bibfnamefont {D.~J.}\ \bibnamefont
  {Luitz}}, \bibinfo {author} {\bibfnamefont {N.}~\bibnamefont {Laflorencie}},\
  and\ \bibinfo {author} {\bibfnamefont {F.}~\bibnamefont {Alet}},\ }\bibfield
  {title} {\bibinfo {title} {Extended slow dynamical regime close to the
  many-body localization transition},\ }\href
  {https://doi.org/10.1103/PhysRevB.93.060201} {\bibfield  {journal} {\bibinfo
  {journal} {Phys. Rev. B}\ }\textbf {\bibinfo {volume} {93}},\ \bibinfo
  {pages} {060201} (\bibinfo {year} {2016})}\BibitemShut {NoStop}%
\bibitem [{\citenamefont {Suntajs}\ \emph
  {et~al.}(2020{\natexlab{a}})\citenamefont {Suntajs}, \citenamefont {Bonca},
  \citenamefont {Prosen},\ and\ \citenamefont {Vidmar}}]{SuntajsBonca}%
  \BibitemOpen
  \bibfield  {author} {\bibinfo {author} {\bibfnamefont {J.}~\bibnamefont
  {Suntajs}}, \bibinfo {author} {\bibfnamefont {J.}~\bibnamefont {Bonca}},
  \bibinfo {author} {\bibfnamefont {T.}~\bibnamefont {Prosen}},\ and\ \bibinfo
  {author} {\bibfnamefont {L.}~\bibnamefont {Vidmar}},\ }\bibfield  {title}
  {\bibinfo {title} {Quantum chaos challenges many-body localization},\ }\href
  {https://doi.org/10.1103/PhysRevE.102.062144} {\bibfield  {journal} {\bibinfo
   {journal} {Phys. Rev. E}\ }\textbf {\bibinfo {volume} {102}},\ \bibinfo
  {pages} {062144} (\bibinfo {year} {2020}{\natexlab{a}})}\BibitemShut
  {NoStop}%
\bibitem [{\citenamefont {Suntajs}\ \emph
  {et~al.}(2020{\natexlab{b}})\citenamefont {Suntajs}, \citenamefont {Bonca},
  \citenamefont {Prosen},\ and\ \citenamefont {Vidmar}}]{SuntajsBonca2}%
  \BibitemOpen
  \bibfield  {author} {\bibinfo {author} {\bibfnamefont {J.}~\bibnamefont
  {Suntajs}}, \bibinfo {author} {\bibfnamefont {J.}~\bibnamefont {Bonca}},
  \bibinfo {author} {\bibfnamefont {T.}~\bibnamefont {Prosen}},\ and\ \bibinfo
  {author} {\bibfnamefont {L.}~\bibnamefont {Vidmar}},\ }\bibfield  {title}
  {\bibinfo {title} {Ergodicity breaking transition in finite disordered spin
  chains},\ }\href {https://doi.org/10.1103/PhysRevB.102.064207} {\bibfield
  {journal} {\bibinfo  {journal} {Phys. Rev. B}\ }\textbf {\bibinfo {volume}
  {102}},\ \bibinfo {pages} {064207} (\bibinfo {year}
  {2020}{\natexlab{b}})}\BibitemShut {NoStop}%
\bibitem [{\citenamefont {Sels}\ and\ \citenamefont
  {Polkovnikov}(2021)}]{SelsPolkovnikov}%
  \BibitemOpen
  \bibfield  {author} {\bibinfo {author} {\bibfnamefont {D.}~\bibnamefont
  {Sels}}\ and\ \bibinfo {author} {\bibfnamefont {A.}~\bibnamefont
  {Polkovnikov}},\ }\bibfield  {title} {\bibinfo {title} {Dynamical obstruction
  to localization in a disordered spin chain},\ }\href
  {https://doi.org/10.1103/PhysRevE.104.054105} {\bibfield  {journal} {\bibinfo
   {journal} {Phys. Rev. E}\ }\textbf {\bibinfo {volume} {104}},\ \bibinfo
  {pages} {054105} (\bibinfo {year} {2021})}\BibitemShut {NoStop}%
\bibitem [{\citenamefont {LeBlond}\ \emph {et~al.}(2021)\citenamefont
  {LeBlond}, \citenamefont {Sels}, \citenamefont {Polkovnikov},\ and\
  \citenamefont {Rigol}}]{SelsPolkovnikov2}%
  \BibitemOpen
  \bibfield  {author} {\bibinfo {author} {\bibfnamefont {T.}~\bibnamefont
  {LeBlond}}, \bibinfo {author} {\bibfnamefont {D.}~\bibnamefont {Sels}},
  \bibinfo {author} {\bibfnamefont {A.}~\bibnamefont {Polkovnikov}},\ and\
  \bibinfo {author} {\bibfnamefont {M.}~\bibnamefont {Rigol}},\ }\bibfield
  {title} {\bibinfo {title} {Universality in the onset of quantum chaos in
  many-body systems},\ }\href {https://doi.org/10.1103/PhysRevB.104.L201117}
  {\bibfield  {journal} {\bibinfo  {journal} {Phys. Rev. B}\ }\textbf {\bibinfo
  {volume} {104}},\ \bibinfo {pages} {L201117} (\bibinfo {year}
  {2021})}\BibitemShut {NoStop}%
\bibitem [{\citenamefont {Sels}\ and\ \citenamefont
  {Polkovnikov}(2022)}]{SelsPolkolnikovReview}%
  \BibitemOpen
  \bibfield  {author} {\bibinfo {author} {\bibfnamefont {D.}~\bibnamefont
  {Sels}}\ and\ \bibinfo {author} {\bibfnamefont {A.}~\bibnamefont
  {Polkovnikov}},\ }\bibfield  {title} {\bibinfo {title} {Whither many-body
  localization?},\ }\href {https://doi.org/10.36471/JCCM_January_2023_01}
  {\bibfield  {journal} {\bibinfo  {journal} {Journal Club for Condensed Matter
  Physics}\ } (\bibinfo {year} {2022})}\BibitemShut {NoStop}%
\bibitem [{\citenamefont {Kiefer-Emmanouilidis}\ \emph
  {et~al.}(2020{\natexlab{a}})\citenamefont {Kiefer-Emmanouilidis},
  \citenamefont {Unanyan}, \citenamefont {Sirker},\ and\ \citenamefont
  {Fleischhauer}}]{KieferSciPost2020}%
  \BibitemOpen
  \bibfield  {author} {\bibinfo {author} {\bibfnamefont {M.}~\bibnamefont
  {Kiefer-Emmanouilidis}}, \bibinfo {author} {\bibfnamefont {R.}~\bibnamefont
  {Unanyan}}, \bibinfo {author} {\bibfnamefont {J.}~\bibnamefont {Sirker}},\
  and\ \bibinfo {author} {\bibfnamefont {M.}~\bibnamefont {Fleischhauer}},\
  }\bibfield  {title} {\bibinfo {title} {Bounds on the entanglement entropy by
  the number entropy in non-interacting fermionic systems},\ }\href@noop {}
  {\bibfield  {journal} {\bibinfo  {journal} {SciPost Phys.}\ }\textbf
  {\bibinfo {volume} {8}},\ \bibinfo {pages} {083} (\bibinfo {year}
  {2020}{\natexlab{a}})}\BibitemShut {NoStop}%
\bibitem [{\citenamefont {Kiefer-Emmanouilidis}\ \emph
  {et~al.}(2020{\natexlab{b}})\citenamefont {Kiefer-Emmanouilidis},
  \citenamefont {Unanyan}, \citenamefont {Fleischhauer},\ and\ \citenamefont
  {Sirker}}]{KieferUnanyan2}%
  \BibitemOpen
  \bibfield  {author} {\bibinfo {author} {\bibfnamefont {M.}~\bibnamefont
  {Kiefer-Emmanouilidis}}, \bibinfo {author} {\bibfnamefont {R.}~\bibnamefont
  {Unanyan}}, \bibinfo {author} {\bibfnamefont {M.}~\bibnamefont
  {Fleischhauer}},\ and\ \bibinfo {author} {\bibfnamefont {J.}~\bibnamefont
  {Sirker}},\ }\bibfield  {title} {\bibinfo {title} {Evidence for unbounded
  growth of the number entropy in many-body localized phases},\ }\href
  {https://doi.org/10.1103/PhysRevLett.124.243601} {\bibfield  {journal}
  {\bibinfo  {journal} {Phys. Rev. Lett.}\ }\textbf {\bibinfo {volume} {124}},\
  \bibinfo {pages} {243601} (\bibinfo {year} {2020}{\natexlab{b}})}\BibitemShut
  {NoStop}%
\bibitem [{\citenamefont {Kiefer-Emmanouilidis}\ \emph
  {et~al.}(2021{\natexlab{a}})\citenamefont {Kiefer-Emmanouilidis},
  \citenamefont {Unanyan}, \citenamefont {Fleischhauer},\ and\ \citenamefont
  {Sirker}}]{KieferUnanyan3}%
  \BibitemOpen
  \bibfield  {author} {\bibinfo {author} {\bibfnamefont {M.}~\bibnamefont
  {Kiefer-Emmanouilidis}}, \bibinfo {author} {\bibfnamefont {R.}~\bibnamefont
  {Unanyan}}, \bibinfo {author} {\bibfnamefont {M.}~\bibnamefont
  {Fleischhauer}},\ and\ \bibinfo {author} {\bibfnamefont {J.}~\bibnamefont
  {Sirker}},\ }\bibfield  {title} {\bibinfo {title} {Slow delocalization of
  particles in many-body localized phases},\ }\href
  {https://doi.org/10.1103/PhysRevB.103.024203} {\bibfield  {journal} {\bibinfo
   {journal} {Phys. Rev. B}\ }\textbf {\bibinfo {volume} {103}},\ \bibinfo
  {pages} {024203} (\bibinfo {year} {2021}{\natexlab{a}})}\BibitemShut
  {NoStop}%
\bibitem [{\citenamefont {Kiefer-Emmanouilidis}\ \emph
  {et~al.}(2021{\natexlab{b}})\citenamefont {Kiefer-Emmanouilidis},
  \citenamefont {Unanyan}, \citenamefont {Fleischhauer},\ and\ \citenamefont
  {Sirker}}]{KieferUnanyan4}%
  \BibitemOpen
  \bibfield  {author} {\bibinfo {author} {\bibfnamefont {M.}~\bibnamefont
  {Kiefer-Emmanouilidis}}, \bibinfo {author} {\bibfnamefont {R.}~\bibnamefont
  {Unanyan}}, \bibinfo {author} {\bibfnamefont {M.}~\bibnamefont
  {Fleischhauer}},\ and\ \bibinfo {author} {\bibfnamefont {J.}~\bibnamefont
  {Sirker}},\ }\bibfield  {title} {\bibinfo {title} {Unlimited growth of
  particle fluctuations in many-body localized phases},\ }\href
  {https://doi.org/https://doi.org/10.1016/j.aop.2021.168481} {\bibfield
  {journal} {\bibinfo  {journal} {Annals of Physics}\ }\textbf {\bibinfo
  {volume} {435}},\ \bibinfo {pages} {168481} (\bibinfo {year}
  {2021}{\natexlab{b}})}\BibitemShut {NoStop}%
\bibitem [{\citenamefont {Kiefer-Emmanouilidis}\ \emph
  {et~al.}(2022{\natexlab{a}})\citenamefont {Kiefer-Emmanouilidis},
  \citenamefont {Unanyan}, \citenamefont {Fleischhauer},\ and\ \citenamefont
  {Sirker}}]{KieferSciPost2022}%
  \BibitemOpen
  \bibfield  {author} {\bibinfo {author} {\bibfnamefont {M.}~\bibnamefont
  {Kiefer-Emmanouilidis}}, \bibinfo {author} {\bibfnamefont {R.}~\bibnamefont
  {Unanyan}}, \bibinfo {author} {\bibfnamefont {M.}~\bibnamefont
  {Fleischhauer}},\ and\ \bibinfo {author} {\bibfnamefont {J.}~\bibnamefont
  {Sirker}},\ }\bibfield  {title} {\bibinfo {title} {Particle fluctuations and
  the failure of simple effective models for many-body localized phases},\
  }\href {https://doi.org/10.21468/SciPostPhys.12.1.034} {\bibfield  {journal}
  {\bibinfo  {journal} {SciPost Phys.}\ }\textbf {\bibinfo {volume} {12}},\
  \bibinfo {pages} {034} (\bibinfo {year} {2022}{\natexlab{a}})}\BibitemShut
  {NoStop}%
\bibitem [{\citenamefont {Imbrie}(2016{\natexlab{a}})}]{Imbrie2016}%
  \BibitemOpen
  \bibfield  {author} {\bibinfo {author} {\bibfnamefont {J.~Z.}\ \bibnamefont
  {Imbrie}},\ }\bibfield  {title} {\bibinfo {title} {Diagonalization and
  many-body localization for a disordered quantum spin chain},\ }\href
  {https://doi.org/10.1103/PhysRevLett.117.027201} {\bibfield  {journal}
  {\bibinfo  {journal} {Phys. Rev. Lett.}\ }\textbf {\bibinfo {volume} {117}},\
  \bibinfo {pages} {027201} (\bibinfo {year} {2016}{\natexlab{a}})}\BibitemShut
  {NoStop}%
\bibitem [{\citenamefont {Imbrie}(2016{\natexlab{b}})}]{Imbrie_JSTAT}%
  \BibitemOpen
  \bibfield  {author} {\bibinfo {author} {\bibfnamefont {J.}~\bibnamefont
  {Imbrie}},\ }\bibfield  {title} {\bibinfo {title} {On many-body localization
  for quantum spin chains},\ }\href {https://doi.org/10.1007/s10955-016-1508-x}
  {\bibfield  {journal} {\bibinfo  {journal} {J. Stat. Phys.}\ }\textbf
  {\bibinfo {volume} {163}},\ \bibinfo {pages} {998} (\bibinfo {year}
  {2016}{\natexlab{b}})}\BibitemShut {NoStop}%
\bibitem [{\citenamefont {Imbrie}\ \emph {et~al.}(2017)\citenamefont {Imbrie},
  \citenamefont {Ros},\ and\ \citenamefont
  {Scardicchio}}]{ImbrieRosScardicchio}%
  \BibitemOpen
  \bibfield  {author} {\bibinfo {author} {\bibfnamefont {J.~Z.}\ \bibnamefont
  {Imbrie}}, \bibinfo {author} {\bibfnamefont {V.}~\bibnamefont {Ros}},\ and\
  \bibinfo {author} {\bibfnamefont {A.}~\bibnamefont {Scardicchio}},\
  }\bibfield  {title} {\bibinfo {title} {Local integrals of motion in many-body
  localized systems},\ }\href
  {https://doi.org/https://doi.org/10.1002/andp.201600278} {\bibfield
  {journal} {\bibinfo  {journal} {Annalen der Physik}\ }\textbf {\bibinfo
  {volume} {529}},\ \bibinfo {pages} {1600278} (\bibinfo {year}
  {2017})}\BibitemShut {NoStop}%
\bibitem [{\citenamefont {{De Roeck}}\ \emph {et~al.}(2024)\citenamefont {{De
  Roeck}}, \citenamefont {Giacomin}, \citenamefont {Huveneers},\ and\
  \citenamefont {Prosniak}}]{deRoeckGiacomin}%
  \BibitemOpen
  \bibfield  {author} {\bibinfo {author} {\bibfnamefont {W.}~\bibnamefont {{De
  Roeck}}}, \bibinfo {author} {\bibfnamefont {L.}~\bibnamefont {Giacomin}},
  \bibinfo {author} {\bibfnamefont {F.}~\bibnamefont {Huveneers}},\ and\
  \bibinfo {author} {\bibfnamefont {O.}~\bibnamefont {Prosniak}},\ }\href
  {https://arxiv.org/abs/2408.04338} {\bibinfo {title} {Absence of normal heat
  conduction in strongly disordered interacting quantum chains}} (\bibinfo
  {year} {2024}),\ \Eprint {https://arxiv.org/abs/2408.04338} {arXiv:2408.04338
  [math-ph]} \BibitemShut {NoStop}%
\bibitem [{\citenamefont {Huveneers}(2026)}]{huveneersMBL}%
  \BibitemOpen
  \bibfield  {author} {\bibinfo {author} {\bibfnamefont {F.}~\bibnamefont
  {Huveneers}},\ }\href {https://arxiv.org/abs/2609.39462} {\bibinfo {title}
  {Many-body localization: an introduction for mathematicians}} (\bibinfo
  {year} {2026}),\ \Eprint {https://arxiv.org/abs/2609.39462} {arXiv:2609.39462
  [math-ph]} \BibitemShut {NoStop}%
\bibitem [{\citenamefont {Morningstar}\ \emph {et~al.}(2020)\citenamefont
  {Morningstar}, \citenamefont {Huse},\ and\ \citenamefont
  {Imbrie}}]{MorningstarHuse}%
  \BibitemOpen
  \bibfield  {author} {\bibinfo {author} {\bibfnamefont {A.}~\bibnamefont
  {Morningstar}}, \bibinfo {author} {\bibfnamefont {D.~A.}\ \bibnamefont
  {Huse}},\ and\ \bibinfo {author} {\bibfnamefont {J.~Z.}\ \bibnamefont
  {Imbrie}},\ }\bibfield  {title} {\bibinfo {title} {Many-body localization
  near the critical point},\ }\href
  {https://doi.org/10.1103/PhysRevB.102.125134} {\bibfield  {journal} {\bibinfo
   {journal} {Phys. Rev. B}\ }\textbf {\bibinfo {volume} {102}},\ \bibinfo
  {pages} {125134} (\bibinfo {year} {2020})}\BibitemShut {NoStop}%
\bibitem [{\citenamefont {De~Roeck}\ and\ \citenamefont
  {Huveneers}(2014)}]{DeRoeckHuveneers}%
  \BibitemOpen
  \bibfield  {author} {\bibinfo {author} {\bibfnamefont {W.}~\bibnamefont
  {De~Roeck}}\ and\ \bibinfo {author} {\bibfnamefont {F.}~\bibnamefont
  {Huveneers}},\ }\bibfield  {title} {\bibinfo {title} {Scenario for
  delocalization in translation-invariant systems},\ }\href
  {https://doi.org/10.1103/PhysRevB.90.165137} {\bibfield  {journal} {\bibinfo
  {journal} {Phys. Rev. B}\ }\textbf {\bibinfo {volume} {90}},\ \bibinfo
  {pages} {165137} (\bibinfo {year} {2014})}\BibitemShut {NoStop}%
\bibitem [{\citenamefont {Weisse}\ \emph {et~al.}(2025)\citenamefont {Weisse},
  \citenamefont {Gerstner},\ and\ \citenamefont
  {Sirker}}]{WeisseGerstnerSirker}%
  \BibitemOpen
  \bibfield  {author} {\bibinfo {author} {\bibfnamefont {A.}~\bibnamefont
  {Weisse}}, \bibinfo {author} {\bibfnamefont {R.}~\bibnamefont {Gerstner}},\
  and\ \bibinfo {author} {\bibfnamefont {J.}~\bibnamefont {Sirker}},\
  }\bibfield  {title} {\bibinfo {title} {Operator growth in disordered spin
  chains: Indications for the absence of many-body localization},\ }\href
  {https://doi.org/10.1103/wgss-nt8t} {\bibfield  {journal} {\bibinfo
  {journal} {Phys. Rev. Res.}\ }\textbf {\bibinfo {volume} {7}},\ \bibinfo
  {pages} {033018} (\bibinfo {year} {2025})}\BibitemShut {NoStop}%
\bibitem [{\citenamefont {Sirker}(2026)}]{SirkerObstruction}%
  \BibitemOpen
  \bibfield  {author} {\bibinfo {author} {\bibfnamefont {J.}~\bibnamefont
  {Sirker}},\ }\href {https://arxiv.org/abs/2606.07774} {\bibinfo {title}
  {Length-resolved operator growth and path-entropy obstructions to many-body
  localization}} (\bibinfo {year} {2026}),\ \Eprint
  {https://arxiv.org/abs/2606.07774} {arXiv:2606.07774} \BibitemShut {NoStop}%
\bibitem [{\citenamefont {Bardarson}\ \emph {et~al.}(2012)\citenamefont
  {Bardarson}, \citenamefont {Pollmann},\ and\ \citenamefont
  {Moore}}]{BardarsonPollmann}%
  \BibitemOpen
  \bibfield  {author} {\bibinfo {author} {\bibfnamefont {J.~H.}\ \bibnamefont
  {Bardarson}}, \bibinfo {author} {\bibfnamefont {F.}~\bibnamefont
  {Pollmann}},\ and\ \bibinfo {author} {\bibfnamefont {J.~E.}\ \bibnamefont
  {Moore}},\ }\bibfield  {title} {\bibinfo {title} {Unbounded growth of
  entanglement in models of many-body localization},\ }\href
  {https://doi.org/10.1103/PhysRevLett.109.017202} {\bibfield  {journal}
  {\bibinfo  {journal} {Phys. Rev. Lett.}\ }\textbf {\bibinfo {volume} {109}},\
  \bibinfo {pages} {017202} (\bibinfo {year} {2012})}\BibitemShut {NoStop}%
\bibitem [{\citenamefont {Serbyn}\ \emph {et~al.}(2013)\citenamefont {Serbyn},
  \citenamefont {Papi\ifmmode~\acute{c}\else \'{c}\fi{}},\ and\ \citenamefont
  {Abanin}}]{Serbyn2013}%
  \BibitemOpen
  \bibfield  {author} {\bibinfo {author} {\bibfnamefont {M.}~\bibnamefont
  {Serbyn}}, \bibinfo {author} {\bibfnamefont {Z.}~\bibnamefont
  {Papi\ifmmode~\acute{c}\else \'{c}\fi{}}},\ and\ \bibinfo {author}
  {\bibfnamefont {D.~A.}\ \bibnamefont {Abanin}},\ }\bibfield  {title}
  {\bibinfo {title} {Universal slow growth of entanglement in interacting
  strongly disordered systems},\ }\href
  {https://doi.org/10.1103/PhysRevLett.110.260601} {\bibfield  {journal}
  {\bibinfo  {journal} {Phys. Rev. Lett.}\ }\textbf {\bibinfo {volume} {110}},\
  \bibinfo {pages} {260601} (\bibinfo {year} {2013})}\BibitemShut {NoStop}%
\bibitem [{\citenamefont {Monkman}\ and\ \citenamefont
  {Sirker}(2023)}]{MonkmanSirker4}%
  \BibitemOpen
  \bibfield  {author} {\bibinfo {author} {\bibfnamefont {K.}~\bibnamefont
  {Monkman}}\ and\ \bibinfo {author} {\bibfnamefont {J.}~\bibnamefont
  {Sirker}},\ }\bibfield  {title} {\bibinfo {title} {Symmetry-resolved
  entanglement: general considerations, calculation from correlation functions,
  and bounds for symmetry-protected topological phases},\ }\href
  {https://doi.org/10.1088/1751-8121/ad086d} {\bibfield  {journal} {\bibinfo
  {journal} {Journal of Physics A: Mathematical and Theoretical}\ }\textbf
  {\bibinfo {volume} {56}},\ \bibinfo {pages} {495001} (\bibinfo {year}
  {2023})}\BibitemShut {NoStop}%
\bibitem [{\citenamefont {Lukin}\ \emph {et~al.}(2019)\citenamefont {Lukin},
  \citenamefont {Rispoli}, \citenamefont {Schittko}, \citenamefont {Tai},
  \citenamefont {Kaufman}, \citenamefont {Choi}, \citenamefont {Khemani},
  \citenamefont {L{\'e}onard},\ and\ \citenamefont
  {Greiner}}]{LukinGreinerMBL}%
  \BibitemOpen
  \bibfield  {author} {\bibinfo {author} {\bibfnamefont {A.}~\bibnamefont
  {Lukin}}, \bibinfo {author} {\bibfnamefont {M.}~\bibnamefont {Rispoli}},
  \bibinfo {author} {\bibfnamefont {R.}~\bibnamefont {Schittko}}, \bibinfo
  {author} {\bibfnamefont {M.~E.}\ \bibnamefont {Tai}}, \bibinfo {author}
  {\bibfnamefont {A.~M.}\ \bibnamefont {Kaufman}}, \bibinfo {author}
  {\bibfnamefont {S.}~\bibnamefont {Choi}}, \bibinfo {author} {\bibfnamefont
  {V.}~\bibnamefont {Khemani}}, \bibinfo {author} {\bibfnamefont
  {J.}~\bibnamefont {L{\'e}onard}},\ and\ \bibinfo {author} {\bibfnamefont
  {M.}~\bibnamefont {Greiner}},\ }\bibfield  {title} {\bibinfo {title} {Probing
  entanglement in a many-body{\textendash}localized system},\ }\href@noop {}
  {\bibfield  {journal} {\bibinfo  {journal} {Science}\ }\textbf {\bibinfo
  {volume} {364}},\ \bibinfo {pages} {256} (\bibinfo {year}
  {2019})}\BibitemShut {NoStop}%
\bibitem [{\citenamefont {Luitz}\ and\ \citenamefont
  {Bar~Lev}(2020)}]{LuitzBarLev}%
  \BibitemOpen
  \bibfield  {author} {\bibinfo {author} {\bibfnamefont {D.~J.}\ \bibnamefont
  {Luitz}}\ and\ \bibinfo {author} {\bibfnamefont {Y.}~\bibnamefont
  {Bar~Lev}},\ }\bibfield  {title} {\bibinfo {title} {Absence of slow particle
  transport in the many-body localized phase},\ }\href
  {https://doi.org/10.1103/PhysRevB.102.100202} {\bibfield  {journal} {\bibinfo
   {journal} {Phys. Rev. B}\ }\textbf {\bibinfo {volume} {102}},\ \bibinfo
  {pages} {100202} (\bibinfo {year} {2020})}\BibitemShut {NoStop}%
\bibitem [{\citenamefont {Ghosh}\ and\ \citenamefont
  {{\v{Z}}nidari{\v{c}}}(2022)}]{GhoshZnidaricPRB2022}%
  \BibitemOpen
  \bibfield  {author} {\bibinfo {author} {\bibfnamefont {R.}~\bibnamefont
  {Ghosh}}\ and\ \bibinfo {author} {\bibfnamefont {M.}~\bibnamefont
  {{\v{Z}}nidari{\v{c}}}},\ }\bibfield  {title} {\bibinfo {title}
  {Resonance-induced growth of number entropy in strongly disordered systems},\
  }\href {https://doi.org/10.1103/PhysRevB.105.144203} {\bibfield  {journal}
  {\bibinfo  {journal} {Phys. Rev. B}\ }\textbf {\bibinfo {volume} {105}},\
  \bibinfo {pages} {144203} (\bibinfo {year} {2022})}\BibitemShut {NoStop}%
\bibitem [{\citenamefont {Aceituno~Ch\'avez}\ \emph {et~al.}(2024)\citenamefont
  {Aceituno~Ch\'avez}, \citenamefont {Artiaco}, \citenamefont {Klein~Kvorning},
  \citenamefont {Herviou},\ and\ \citenamefont
  {Bardarson}}]{AceitunoChavezPRL2024}%
  \BibitemOpen
  \bibfield  {author} {\bibinfo {author} {\bibfnamefont {D.}~\bibnamefont
  {Aceituno~Ch\'avez}}, \bibinfo {author} {\bibfnamefont {C.}~\bibnamefont
  {Artiaco}}, \bibinfo {author} {\bibfnamefont {T.}~\bibnamefont
  {Klein~Kvorning}}, \bibinfo {author} {\bibfnamefont {L.}~\bibnamefont
  {Herviou}},\ and\ \bibinfo {author} {\bibfnamefont {J.~H.}\ \bibnamefont
  {Bardarson}},\ }\bibfield  {title} {\bibinfo {title} {Ultraslow growth of
  number entropy in an $\ell$-bit model of many-body localization},\ }\href
  {https://doi.org/10.1103/PhysRevLett.133.126502} {\bibfield  {journal}
  {\bibinfo  {journal} {Phys. Rev. Lett.}\ }\textbf {\bibinfo {volume} {133}},\
  \bibinfo {pages} {126502} (\bibinfo {year} {2024})}\BibitemShut {NoStop}%
\bibitem [{\citenamefont {Faulend}\ \emph {et~al.}(2026)\citenamefont
  {Faulend}, \citenamefont {Buljan},\ and\ \citenamefont
  {Štrkalj}}]{faulend2026}%
  \BibitemOpen
  \bibfield  {author} {\bibinfo {author} {\bibfnamefont {B.}~\bibnamefont
  {Faulend}}, \bibinfo {author} {\bibfnamefont {H.}~\bibnamefont {Buljan}},\
  and\ \bibinfo {author} {\bibfnamefont {A.}~\bibnamefont {Štrkalj}},\ }\href
  {https://arxiv.org/abs/2603.28721} {\bibinfo {title} {Quantum beats in
  many-body localized systems}} (\bibinfo {year} {2026}),\ \Eprint
  {https://arxiv.org/abs/2603.28721} {arXiv:2603.28721 [cond-mat.dis-nn]}
  \BibitemShut {NoStop}%
\bibitem [{\citenamefont {Kiefer-Emmanouilidis}\ \emph
  {et~al.}(2022{\natexlab{b}})\citenamefont {Kiefer-Emmanouilidis},
  \citenamefont {Unanyan}, \citenamefont {Fleischhauer},\ and\ \citenamefont
  {Sirker}}]{KieferComment2203}%
  \BibitemOpen
  \bibfield  {author} {\bibinfo {author} {\bibfnamefont {M.}~\bibnamefont
  {Kiefer-Emmanouilidis}}, \bibinfo {author} {\bibfnamefont {R.}~\bibnamefont
  {Unanyan}}, \bibinfo {author} {\bibfnamefont {M.}~\bibnamefont
  {Fleischhauer}},\ and\ \bibinfo {author} {\bibfnamefont {J.}~\bibnamefont
  {Sirker}},\ }\href@noop {} {\bibinfo {title} {Comment on ``resonance-induced
  growth of number entropy in strongly disordered systems''}} (\bibinfo {year}
  {2022}{\natexlab{b}}),\ \Eprint {https://arxiv.org/abs/2203.06689}
  {arXiv:2203.06689} \BibitemShut {NoStop}%
\bibitem [{\citenamefont {Vu}\ \emph {et~al.}(2022)\citenamefont {Vu},
  \citenamefont {Huang}, \citenamefont {Li},\ and\ \citenamefont
  {Das~Sarma}}]{VuHuang}%
  \BibitemOpen
  \bibfield  {author} {\bibinfo {author} {\bibfnamefont {D.}~\bibnamefont
  {Vu}}, \bibinfo {author} {\bibfnamefont {K.}~\bibnamefont {Huang}}, \bibinfo
  {author} {\bibfnamefont {X.}~\bibnamefont {Li}},\ and\ \bibinfo {author}
  {\bibfnamefont {S.}~\bibnamefont {Das~Sarma}},\ }\bibfield  {title} {\bibinfo
  {title} {Fermionic many-body localization for random and quasiperiodic
  systems in the presence of short- and long-range interactions},\ }\href
  {https://doi.org/10.1103/PhysRevLett.128.146601} {\bibfield  {journal}
  {\bibinfo  {journal} {Phys. Rev. Lett.}\ }\textbf {\bibinfo {volume} {128}},\
  \bibinfo {pages} {146601} (\bibinfo {year} {2022})}\BibitemShut {NoStop}%
\bibitem [{\citenamefont {Tu}\ \emph {et~al.}(2023)\citenamefont {Tu},
  \citenamefont {Vu},\ and\ \citenamefont {Das~Sarma}}]{TuDasSarma2023}%
  \BibitemOpen
  \bibfield  {author} {\bibinfo {author} {\bibfnamefont {Y.-T.}\ \bibnamefont
  {Tu}}, \bibinfo {author} {\bibfnamefont {D.}~\bibnamefont {Vu}},\ and\
  \bibinfo {author} {\bibfnamefont {S.}~\bibnamefont {Das~Sarma}},\ }\bibfield
  {title} {\bibinfo {title} {Avalanche stability transition in interacting
  quasiperiodic systems},\ }\href@noop {} {\bibfield  {journal} {\bibinfo
  {journal} {Phys. Rev. B}\ }\textbf {\bibinfo {volume} {107}},\ \bibinfo
  {pages} {014203} (\bibinfo {year} {2023})}\BibitemShut {NoStop}%
\bibitem [{\citenamefont {Padhan}\ \emph {et~al.}(2026)\citenamefont {Padhan},
  \citenamefont {Colbois}, \citenamefont {Alet},\ and\ \citenamefont
  {Laflorencie}}]{PadhanColbois}%
  \BibitemOpen
  \bibfield  {author} {\bibinfo {author} {\bibfnamefont {A.}~\bibnamefont
  {Padhan}}, \bibinfo {author} {\bibfnamefont {J.}~\bibnamefont {Colbois}},
  \bibinfo {author} {\bibfnamefont {F.}~\bibnamefont {Alet}},\ and\ \bibinfo
  {author} {\bibfnamefont {N.}~\bibnamefont {Laflorencie}},\ }\bibfield
  {title} {\bibinfo {title} {Long-range resonances in quasiperiodic many-body
  localization},\ }\href {https://doi.org/10.1103/h9mp-t4qz} {\bibfield
  {journal} {\bibinfo  {journal} {Phys. Rev. Lett.}\ }\textbf {\bibinfo
  {volume} {136}},\ \bibinfo {pages} {197101} (\bibinfo {year}
  {2026})}\BibitemShut {NoStop}%
\bibitem [{\citenamefont {Jitomirskaya}(1999)}]{Jitomirskaya1999}%
  \BibitemOpen
  \bibfield  {author} {\bibinfo {author} {\bibfnamefont {S.~Y.}\ \bibnamefont
  {Jitomirskaya}},\ }\bibfield  {title} {\bibinfo {title} {Metal-insulator
  transition for the almost mathieu operator},\ }\href
  {https://doi.org/doi.org/10.2307/121066} {\bibfield  {journal} {\bibinfo
  {journal} {Ann. Math.}\ }\textbf {\bibinfo {volume} {150}},\ \bibinfo {pages}
  {1159} (\bibinfo {year} {1999})}\BibitemShut {NoStop}%
\end{thebibliography}
%apsrev4-2.bst 2019-01-14 (MD) hand-edited version of apsrev4-1.bst
%Control: key (0)
%Control: author (8) initials jnrlst
%Control: editor formatted (1) identically to author
%Control: production of article title (0) allowed
%Control: page (0) single
%Control: year (1) truncated
%Control: production of eprint (0) enabled
%

\end{document}